\documentclass[pdflatex,sn-mathphys,Numbered]{sn-jnl}
\usepackage{a4wide}
\usepackage{mathpazo}
\usepackage{graphicx}%
\usepackage{multirow}%
\usepackage{amsmath,amssymb,amsfonts}%
\usepackage{amsthm}%
\usepackage{mathrsfs}%
\usepackage[title]{appendix}%
\usepackage{xcolor}%
\usepackage{textcomp}%
\usepackage{manyfoot}%
\usepackage{booktabs}%
\usepackage{algorithm}%
\usepackage{algorithmicx}%
\usepackage{algpseudocode}%
\usepackage{listings}%
\usepackage{adjustbox}
\usepackage[normalem]{ulem}
\usepackage{paracol}
\begin{document}
\title{
Food co-purchasing networks encode geographic disparities in cardiometabolic health
}
\author[a,b,c,$*$]{\fnm{Antonio} \sur{Desiderio}}
\author[a]{\fnm{Leonardo} \sur{Rodovero}}
\author[a,d,$*$]{\fnm{Sune} \sur{Lehmann}}
\affil[a]{\orgdiv{Department of Applied Mathematics and Computer Science}, \orgname{Technical University of Denmark}, \orgaddress{\street{Richard Petersens Plads}, \postcode{2800}, \state{Copenhagen}, \country{Denmark}}}
\affil[b]{\orgdiv{ISI Foundation}, \orgaddress{\street{Via Chisola 5}, \postcode{10126}, \state{Turin}, \country{Italy}}}
\affil[c]{\orgname{Centro Ricerche Enrico Fermi}, \orgaddress{\street{Via Panisperna 89a}, \postcode{00184}, \state{Rome}, \country{Italy}}}
\affil[d]{\orgdiv{Copenhagen Center for Social Data Science (SODAS)}, \orgname{University of Copenhagen}, \orgaddress{\street{Øster Farimagsgade 5}, \postcode{1353}, \state{Copenhagen}, \country{Denmark}}}
\affil[$*$]{Corresponding authors: \href{mailto:antde@dtu.dk}{antde@dtu.dk} and \href{mailto:sljo@dtu.dk}{sljo@dtu.dk}}
\abstract{
Unhealthy diets are a major driver of cardiometabolic disease, yet the geospatial structure of dietary behavior and its links to population health remain poorly understood.
While large-scale retail transaction data offer a unique view of population dietary behavior, conventional approaches reduce purchases to isolated items, not accounting for how foods are actually combined.
Here, we analyze 51 million grocery receipts across 95 Danish municipalities using a statistically validated bipartite network framework to isolate co-purchases that product popularity alone cannot explain.
We find that, although municipalities buy largely the same products, their co-purchasing networks reorganize along a continuous urban-to-peripheral gradient in how those products are combined.
The network backbone can be resolved into modular dietary communities whose local cohesion allows us to predict cardiometabolic prescriptions, explaining a quarter to a half of the variance left after sociodemographic adjustment, beyond what product prevalence alone accounts for.
Ultra-processed foods (UPFs) show no spatial gradient in prevalence and are located at the network periphery, where they share baskets with fresh foods but rarely form the combinations that hold the network together.
Taken together, our findings show that geographic dietary differences are encoded not only in what people purchase, but in how foods are systematically combined within shopping baskets, offering a scalable network approach for population nutritional epidemiology.
}
\maketitle
\section*{Introduction}
Non-communicable diseases, including cardiovascular disease and type 2 diabetes, impose a major global health burden \cite{zhou_worldwide_2016,afshin_health_2019}, and unhealthy dietary patterns contribute significantly to their incidence and mortality \cite{jayedi_healthy_2020,ohearn_incident_2023,micha_red_2010,wang_optimal_2023,huang_dietary_2023,lane_ultra-processed_2024,springmann_analysis_2016,de_souza_intake_2015,tilman_global_2014}.
Food-system transformation \cite{popkin_nutrition_2001,bilal_neighborhood_2018,muncke_health_2025} -- driven by industrialization, urbanization, and the expansion of retail chains \cite{reardon_economics_2012,seto_hidden_2016,ambikapathi_global_2022,di_clemente_urbanization_2021} -- has reconfigured the availability and affordability of food across the past decades \cite{schneider_state_2023,scapin_global_2025,popkin_towards_2021}, while the growing market penetration and widespread consumption of ultra-processed foods \cite{monteiro_ultra-processed_2019,monteiro_decade_2018,scrinis_ultra-processed_2022,vandevijvere_global_2019,coyle_socio-economic_2022,scapin_global_2025,martinez_steele_ultra-processed_2016,the_lancet_ultra-processed_2025} have changed individual dietary profiles, in a society-wide shift often associated with adverse cardiometabolic outcomes \cite{elizabeth_ultra-processed_2020,hall_ultra-processed_2019,lane_ultra-processed_2024,monteiro_ultra-processed_2025,bonaccio_ultra-processed_2021,cordova_consumption_2023,srour_ultra-processed_2022,juul_role_2025,oneill_ultra-processed_2026}.
These health risks are not distributed uniformly across society; there is geographical variation in dietary patterns that reflects spatial and socioeconomic inequalities \cite{ncd_risk_factor_collaborationncd-risc_rising_2019,he_healthenvironment_2024,allcott_food_2017,zheng_health_2024}, rooted in underlying structural disparities in local food environments \cite{cummins_food_2006,franco_availability_2009,moore_fast-food_2009,xu_integrating_2023,althoff_large-scale_2022,popkin_nutrition_2001,desiderio_mapping_2026}.

Measuring how diets vary across places, however, is difficult.
Traditional nutritional epidemiology relies on dietary surveys and longitudinal cohorts, which provide detailed information on food consumption patterns but are constrained by self-reporting biases, high implementation costs, and limited spatial coverage \cite{afshin_health_2019,fiolet_consumption_2018}.
Importantly, this tradition has long recognized that foods are eaten in combination, as dietary-pattern analyses capture how individuals combine foods, and these patterns reveal diet--disease relationships that single foods or nutrients miss \cite{hu_dietary_2002, iqbal_gaussian_2016, jayedi_healthy_2020}.

Computational methods now offer a way around the constraints of surveys, opening up a data-driven nutritional epidemiology \cite{barabasi_unmapped_2019}.
This field uses large-scale digital records, such as retail transaction data \cite{bandy_use_2019,nevalainen_large-scale_2018,pennacchioli_retail_2014,aiello_large-scale_2019,althoff_large-scale_2022,ravandi_prevalence_2025,fox_consumer_2004}, to trace the complex system of human interactions with food at a level of detail that was previously out of reach \cite{rutter_need_2017}; such records provide unprecedented opportunities to quantify dietary habits across entire populations \cite{ni_mhurchu_monitoring_2013,kuhl_ai_2025}.
Yet these rich transaction datasets are almost always reduced to product- or category-level summaries \cite{nevalainen_large-scale_2018,aiello_large-scale_2019,bandy_use_2019,martin_feasibility_2006,pennacchioli_retail_2014,althoff_large-scale_2022}, discarding exactly the combination information that dietary-pattern research has shown to matter \cite{mozaffarian_history_2018,menichetti_machine_2023}.

Bringing the dietary-pattern perspective to transaction data -- understanding how the network organization of food co-purchases varies across geographic contexts, and whether this relational structure encodes population health signals beyond individual product prevalence -- is therefore the natural next step in nutritional modeling, and a new way to identify drivers of nutritional disparities \cite{johansen_food_2006,pennacchioli_retail_2014,aiello_large-scale_2019,dong_social_2018,dong_predicting_2019}.

Here, we characterize population dietary behavior through food co-purchasing patterns, and ask what food combinations reveal that individual products cannot \cite{johansen_food_2006,mozaffarian_history_2018}.
By examining how foods are combined rather than considering individual products in isolation, we identify a layer of dietary structure that varies across places and tracks population health.
Specifically, we develop a network-based framework to identify non-random co-purchasing relationships and examine how their organization varies across geographic contexts \cite{saracco_inferring_2017, saracco_randomizing_2015, pennacchioli_retail_2014, cimini_statistical_2019}.
We then assess how these network structures capture population-level cardiometabolic health signals beyond individual product prevalence and demographic composition \cite{hunter_social_2017,dong_predicting_2019,kuhl_ai_2025}.
We apply our approach to large-scale grocery purchasing data from Denmark, where eating out is relatively rare: about 85\% of consumers describe their most recent dinner as home-prepared, leftovers, or based on ready-made components \cite{lf_danskernes_aftensmad_2024}, so grocery purchases capture a substantial share of the food the population eats.
We find that co-purchasing networks reorganize along a continuous urban-to-peripheral gradient, and that dietary community cohesion tracks municipal cardiometabolic prescriptions beyond sociodemographics and local product prevalence.
Notably, while ultra-processed food sales show no geographic variation, these products consistently sit at the network periphery.
\section*{Results}

\textbf{Every receipt from a nationwide supermarket chain.}
To investigate the spatial organization of dietary habits and its relationship with population-level cardiometabolic health, we draw on retail purchasing records combined with national nutrition databases \cite{poulsen_dataset_2026} and official public-health registries in Denmark \cite{statistics_denmark_statbank_2026}.
Through a collaboration with Salling Group (Denmark's largest retailer), we analyzed 51 million point-of-sale receipts from 553 Netto supermarket locations across 95 of Denmark's 98 municipalities, collected over 79 days in 2025 (see Methods).
To isolate dietary purchases, non-food transactions (carrier bags, personal care, household goods, seasonal non-food merchandise, and related categories) were removed, and edible items were harmonized at the brand-and-microcategory level (e.g. ``Brand of Milk'' -- whole milk), yielding 3,405 ``products'' after exclusion (see Methods and Supplementary Materials, Section 1).

\textbf{Receipt data linked to nutritional and processing measures.}
We matched each food product to a reference entry in the Danish national food-composition database (Frida) \cite{poulsen_dataset_2026} using a hierarchical multi-stage semantic matching approach (see Methods for the complete pipeline).
This mapping establishes both a nutritional and an industrial processing profile for each item, capturing energy-adjusted macronutrients alongside a continuous ultra-processed food (UPF) risk index ($F\text{Score} \in [0, 1]$) and categorical NOVA classes \cite{monteiro_ultra-processed_2019,monteiro_decade_2018}, estimated via machine learning (FoodProX) \cite{menichetti_machine_2023} and benchmarked against Open Food Facts records (see Methods for further details).

To evaluate population-level outcomes, we aggregated these item-level metrics and aligned the resulting municipal grocery profiles with 2024 national health registries from Statistics Denmark across the 95 municipalities with retail coverage \cite{statistics_denmark_statbank_2026}.
The cardiometabolic burden for a municipality was measured as the share of residents who redeemed at least one prescription in 2024 for antidiabetic medicines (ATC A10), cardiovascular-system medicines (ATC C), or lipid-modifying agents (ATC C10).
We note that these rates are treated-disease proxies rather than clinical diagnoses; for instance, A10 combines insulin and non-insulin therapies, effectively capturing treated diabetes as a whole -- roughly 93\% of which is type 2 (estimated from regional rather than municipality-level data; see Methods).
Finally, we linked municipal grocery and health series to Statistics Denmark's municipality-level data for population, age, and disposable family income (see Methods for further details).
\subsection*{The geography of how foods are co-purchased}
\begin{figure*}[!htbp]
    \centering\includegraphics[width=0.895\textwidth]{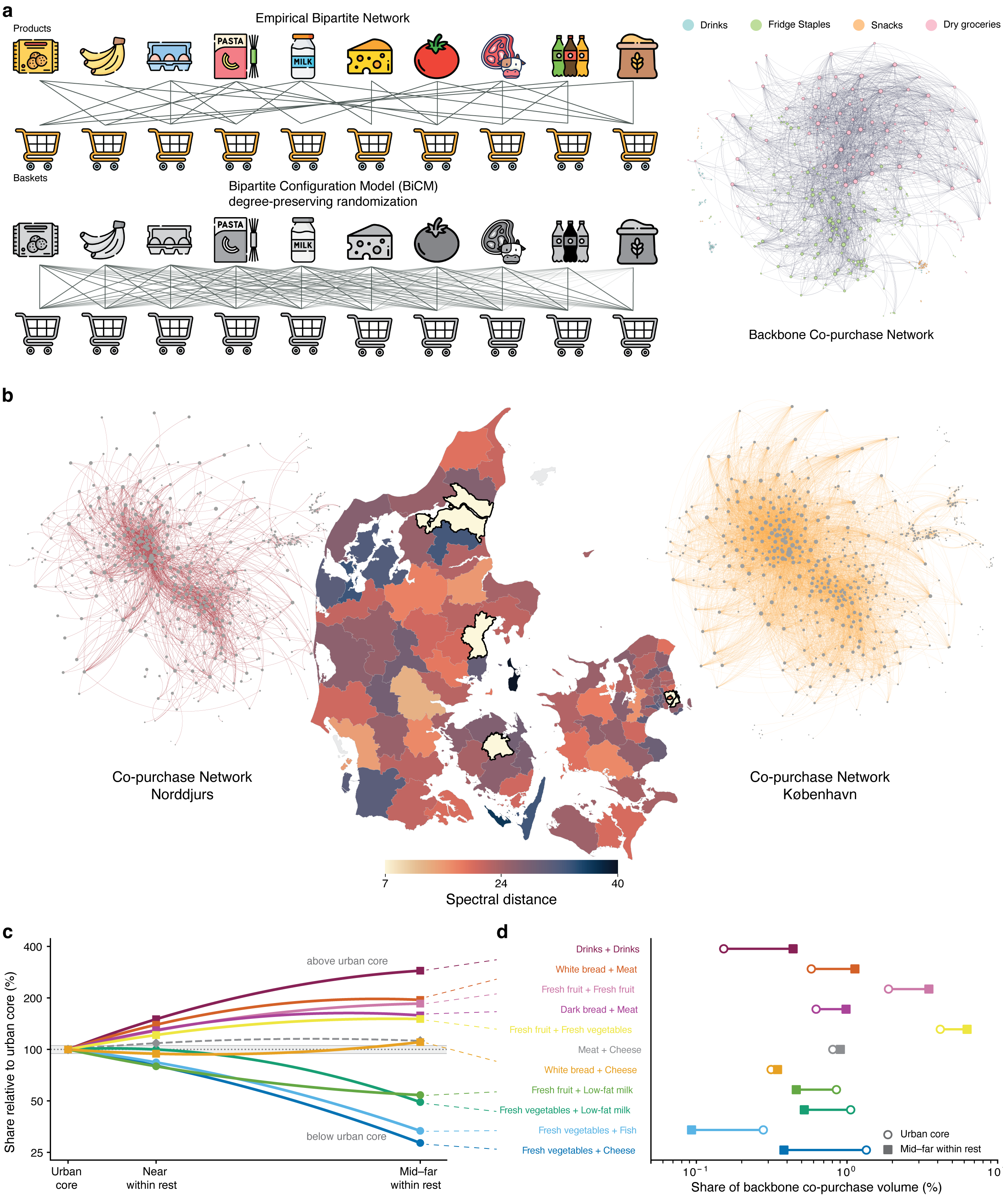}
    \caption{\textbf{Statistical validation and spatial reorganization of food co-purchasing networks across Denmark.}
    \textbf{a)} Empirical basket--product records (top left) are compared with a Bipartite Configuration Model (BiCM, bottom left; edges are shaded proportionally to connection probabilities under the null model), which preserves observed basket sizes and product degrees. Statistically validated associations are projected onto a monopartite product--product network -- representing the Randers co-purchasing network (right, nodes colored by broad product group) -- retaining co-purchasing relationships occurring more often than expected under the BiCM.
    \textbf{b)} Danish municipalities colored by their spectral distance from the four-city urban core (outlined), based on the normalized-Laplacian spectra of their validated networks; municipalities without data are gray. Norddjurs (left) and K{\o}benhavn (right) are shown with identical node positions, illustrating topological rewiring despite a largely shared product set.
    On this shared product set, normalized degree entropy is 0.66 in K{\o}benhavn and 0.39 in Norddjurs, and the random-walk entropy rate is 4.30 and 2.96; this topological contrast persists under equalized sample sizes, basket-size matching, and effect-size filtering (see Supplementary Materials, Section 2).
    \textbf{c)} Mean share of backbone co-purchase volume for selected meta-category pairs, relative to the urban-core mean ($100\% = \text{parity}$) across the urban core, near, and mid--far groups; pair colors as in \textbf{d}. Warm and cool colors indicate increases and decreases away from the urban core, respectively. Points show group means; curves are $\log_2$-space interpolations. The gray band denotes the largest standard error of the mean of the meat--cheese control pair.
    \textbf{d)} Absolute mean share of backbone co-purchase volume for selected pairs in the urban core (open circles) and mid--far municipalities (filled squares), highlighting a peripheral shift toward staple and within-category pairings alongside a decline in fresh food--dairy/fish combinations.
    }
\label{fig_1}
\end{figure*}

\textbf{Co-occurrences beyond product popularity.}
Raw co-occurrence frequencies \cite{pennacchioli_retail_2014,tufano_data-driven_2025,johansen_food_2006,aiello_large-scale_2019} can confound genuine dietary affinities because of the heavy-tailed distribution of product popularity: ubiquitous staples such as milk or bread will tend to co-occur frequently with a wide range of items simply because their high purchase rate inflates the baseline probability of random overlaps, rather than because they reflect specific dietary complementarity.
To filter out these popularity artifacts, we validate basket-level co-purchasing against the Bipartite Configuration Model (BiCM) \cite{saracco_randomizing_2015,saracco_inferring_2017,cimini_statistical_2019}, representing transactions as a bipartite network of shopping baskets and food products and preserving observed marginal degrees -- namely, basket sizes and product purchase frequencies (see Figure \ref{fig_1}\textbf{a} and Methods).
The resulting validated backbone retains only product pairs that co-occur significantly more often than expected by chance.
We extract this backbone for each Danish municipality, allowing regional dietary structures to emerge from the way foods are combined in actual purchases rather than from differences in overall product popularity or purchasing volume.

\textbf{While products are broadly shared, foods are combined differently across the country.}
Using our co-purchasing backbone for each municipality, we first examine whether we find a uniform `national diet' or distinct local patterns of food associations.
We compare the overall wiring of the municipal networks using the spectral distance between their normalized Laplacians, which captures differences in network structure independently of geographic distance (see Methods for further details).
Denmark’s four largest cities -- K{\o}benhavn, Aarhus, Odense and Aalborg -- stand out as a distinct cluster, which we call the \emph{urban core} (pairwise spectral distances below $10$, against a median of $25$ among the remaining municipalities; hierarchical clustering isolates the same four-city block).
Clustering on shared edges instead recovers the same partition (Jaccard distance between validated edge sets; adjusted Rand index $0.97$).
Conversely, municipal product composition is highly uniform across the country (median Jaccard similarity $=0.65$) and does not isolate the four-city cluster, yielding no clear geographic partitions beyond a few small-catalog outliers.

Measuring each municipality's mean spectral distance $d$ to the urban core gives us a continuous urban-to-peripheral gradient, mapped in Figure \ref{fig_1}\textbf{b} alongside the networks of K{\o}benhavn and Norddjurs (a non-core municipality with an overlapping catalog), drawn with identical node coordinates to highlight differences in edge wiring.
For instance, across $\sim$1,000 common products, K{\o}benhavn shows a more heterogeneous co-purchasing topology (degree entropy 0.66 vs. 0.39; random-walk entropy rate 4.30 vs. 2.96), a contrast that persists under equalized sample sizes and matched basket-size distributions.
Holding out the urban core, we partition the remaining 91 municipalities at the median distance ($d=25.1$) into \emph{near} ($n=47$, median $d=22$) and \emph{mid--far} ($n=44$, median $d=29$) groups.
Along this gradient, edge Jaccard similarity with the urban core drops monotonically from 0.55 within the core to 0.25 in the near group and 0.12 in the mid--far group (Kruskal--Wallis test $H=70.7$, $p < 0.001$; Mann--Whitney test near vs. mid--far, $p< 0.001$).
Recomputing spectral distances strictly on pairwise shared products without spectrum padding preserves this urban-to-peripheral ordering ($\rho = 0.78$ pairwise and $\rho = 0.85$ to the urban core), and the geographic patterns remain stable across per-capita sampling ratios and random sampling seeds (see Methods and Supplementary Materials, Section 2).
Per-capita sampling retains population scale as part of the ecological description of municipal purchasing activity, although spectral distance remains strongly associated with sample size ($\rho = -0.975$) and validated-catalog size ($\rho = -0.999$).
To isolate basket composition from sample size under Bonferroni validation, we first compared K{\o}benhavn and Norddjurs directly at equal sample sizes, under a size-invariant effect threshold ($z/\sqrt{N}$), and under permutations matched on basket size.
Across all three controls, the spectral distance between K{\o}benhavn and Norddjurs stays well above the label-permutation null ($p < 0.001$), with Odense at closer distance and neighboring Frederiksberg indistinguishable from K{\o}benhavn ($p \ge 0.10$).
Extending this equal-sample, basket-size-matched design to all 95 municipalities ($N^{*} = 53{,}251$ baskets) consolidates 21 of the 27 smaller Capital Region suburbs into the closest tertile around K{\o}benhavn, while the largest regional cities remain distinct from their rural surroundings.
This suburban consolidation compresses edge overlap when evaluated on the unadjusted population-scaled groups ($0.47$, $0.45$, and $0.43$), yet municipal rankings outside the Capital Region remain consistent with the per-capita gradient ($\rho = 0.61$, $p < 10^{-7}$) and edge overlap with K{\o}benhavn declines monotonically across fixed-$N^{*}$ tertiles ($0.44$, $0.40$, and $0.36$; $\rho = -0.85$ nationwide and $\rho = -0.80$ outside the Capital Region; see Supplementary Materials, Section 2).

\textbf{Shifts in food co-occurrence patterns across the urban--peripheral gradient.}
Having established that co-purchasing networks are reorganized along a spectral urban-to-peripheral gradient despite largely shared product catalogs, we next ask which food combinations carry this rewiring.
For each municipality, we take the backbone's co-purchase volume -- the number of baskets containing each validated pair, summed over all pairs -- and compute the share falling on each pair of meta-categories.
We then track each pair's mean share relative to the urban core along spectral distance (see Figure~\ref{fig_1}\textbf{c} and \textbf{d}).
Pairings between fresh foods and dairy or fish, which are more prominent in the urban core, decline towards the mid--far municipalities, with vegetables--cheese falling from 1.34\% to 0.38\% of the backbone (28\% of the urban-core share; $\rho=-0.89$, $p<0.001$), vegetables--fish reaching 33\% of the urban-core share ($\rho=-0.50$, $p<0.001$), and vegetables or fruit with low-fat milk to 49--54\% ($\rho=-0.58$, $p<0.001$, and $\rho=-0.23$, $p=0.025$).
Conversely, fruit--vegetable rises from 4.16\% to 6.27\% (151\% of the urban-core share) and fruit--fruit from 1.89\% to 3.49\% (185\%; $\rho=0.59$ and $0.76$; both $p<0.001$), while white bread--meat reaches 195\% of the urban-core share ($\rho=0.42$, $p<0.001$).
Drinks--drinks shows the steepest relative increase (290\% of the urban-core share; $\rho=0.61$, $p<0.001$), while remaining a small share of the backbone.
These urban-to-rural shifts in food pairings hold both on unpadded shared-product spectra and when comparing rural municipalities directly to the urban core under fixed-$N^{*}$ sampling (with rural-to-urban ratios of $0.58$ vs.\ $0.59$ for vegetables--cheese, $0.45$ vs.\ $0.30$ for vegetables--fish, $1.84$ vs.\ $1.65$ for white bread--meat, and $2.13$ vs.\ $2.20$ for drinks--drinks; see Supplementary Materials, Section 2).
Thus, despite the nationwide standardization of supermarket supply chains, the ways in which consumers combine products in shopping baskets remain strongly localized.
\subsection*{Food co-purchasing structure is associated with cardiometabolic health}
\textbf{Deriving a robust national food co-purchasing backbone.}
To evaluate whether and how local food purchasing structures correlate with municipal-level differences in cardiometabolic health \cite{aiello_large-scale_2019}, we need a shared national reference space across municipalities.
To establish this framework, we pool population-normalized receipts and apply the bipartite configuration model (BiCM) to construct a national co-purchasing backbone (see Methods).
We then filter the backbone using the BiCM-derived $z$-score at a percolation threshold of $z^{*}=28$, retaining 95\% of products within a single giant connected component (2,659 nodes and 28,435 edges, resulting in a density of $0.008$) \cite{saracco_randomizing_2015,saracco_inferring_2017}.
We choose this threshold to preserve network connectivity while reducing the dense catalog of significant co-purchases; importantly, $z$-score filtering isolates the structural backbone without biasing the underlying community partition (see Supplementary Materials, Section 3).
The resulting sparse network has community structure, with a modularity of $Q=0.48$.
Community detection identifies 12 major coherent dietary communities -- including Ready Meals, Cheese \& cold cuts, Candy \& snacks, Wine \& beer, and Pasta \& toppings.
Crucially, this mesoscale organization is structurally robust.
We find strong agreement between two community detection methods built on different principles, Louvain \cite{blondel_fast_2008} and Infomap \cite{rosvall_maps_2008} (Adjusted Mutual Information \(\approx0.70\)).
A Jaccard-based co-occurrence network \cite{johansen_food_2006, pennacchioli_retail_2014}, however, does not recover this organization, remaining dense ($\sim1.2$ million edges, density $\sim 0.3$) with little modular structure at the median threshold, while progressively stricter thresholds reduce connectivity without revealing a comparable community backbone ($Q\lesssim0.2$).

\begin{figure*}[!htbp]
    \centering\includegraphics[width=1.0\textwidth]{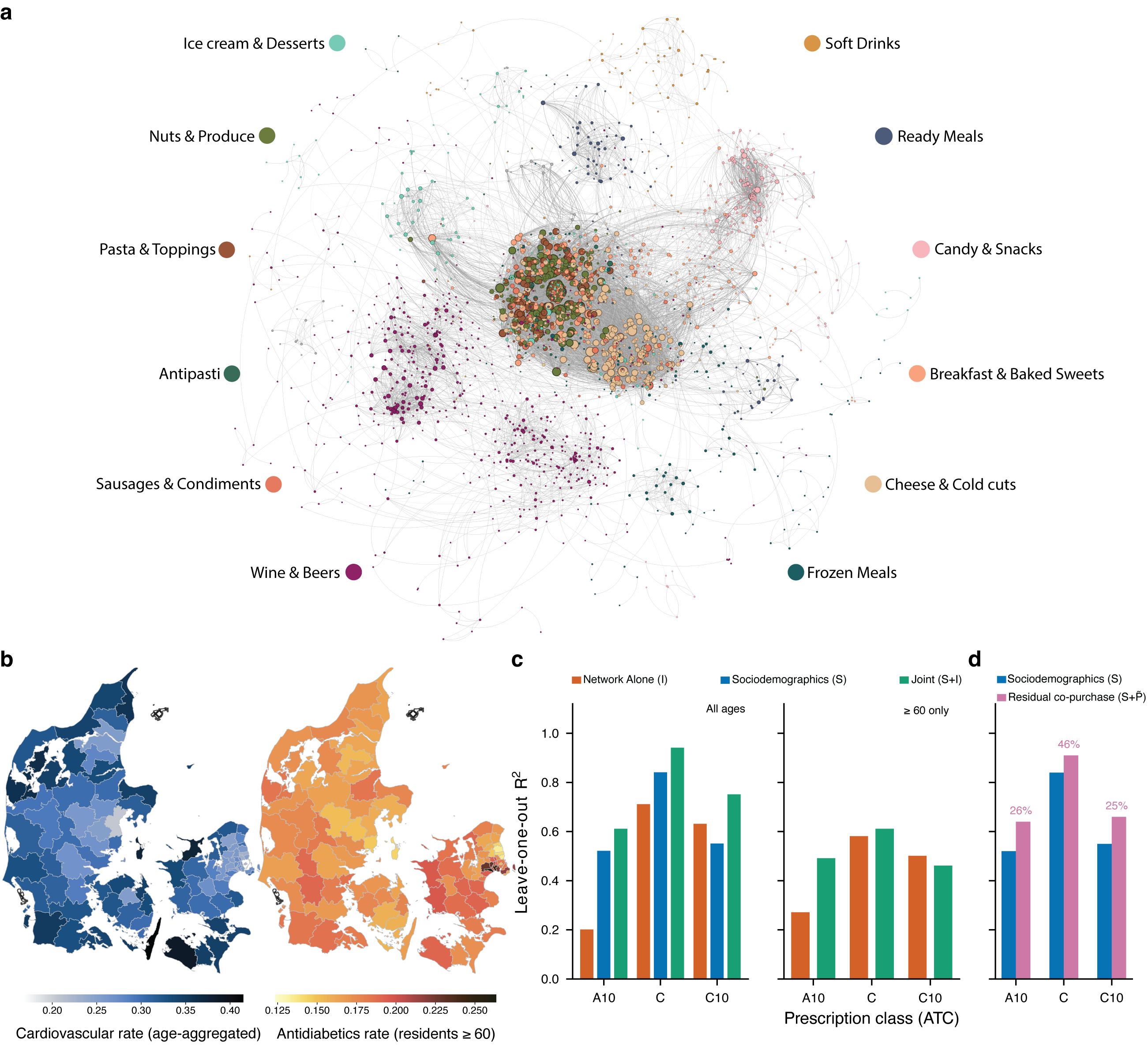}
    \caption{\textbf{Food co-purchasing network and its association with cardiometabolic health disparities across Denmark.}
    \textbf{a)} National validated co-purchasing backbone ($z^* = 28$; 2,659 nodes, 28,435 edges), partitioned into 12 major dietary communities (node colors) with strong modular organization ($Q = 0.48$).
    \textbf{b)} Share of residents redeeming at least one prescription in 2024 for cardiovascular medicines (ATC C, all ages; left) and antidiabetic medicines (ATC A10, residents aged $\ge 60$; right). Fan{\o} and L{\ae}s{\o} are white-hatched (no complete coverage; excluded), Bornholm omitted.
    \textbf{c)} Out-of-sample predictive performance (leave-one-out $R^2$) for antidiabetic (A10), cardiovascular (C), and lipid-modifying (C10) prescription rates, for all ages (left) and residents aged $\ge 60$ (right). Bars compare local network intensity alone ($I$, orange), sociodemographics alone ($S$, blue; omitted for residents aged $\ge 60$ as $R^2 < 0.07$), and both together ($S+I$, green).
    \textbf{d)} Predictive gain from co-purchasing structure beyond product prevalence (all ages). Adding residualized co-purchase cohesion ($S+\widetilde{P}$, pink) to sociodemographics ($S$, blue) improves the fit for all three outcomes, closing 25--46\% of the variance left unexplained by $S$ (percentages above the bars; see Supplementary Materials, Section 4 for the community-level regression coefficients).
    }
\label{fig_2}
\end{figure*}

\textbf{Local community cohesion predicts municipal prescription rates.}
To relate national dietary communities to cardiometabolic burden, we summarize each municipality by the intensity of its locally dense co-purchase communities, measured as the mean intra-community log-lift $I_c$ of the national backbone evaluated on local baskets (see Methods for details).
First, we predict age-aggregated 2024 redemption rates (see Figure \ref{fig_2}\textbf{b}) from the intensity vector $I_c$ alone and jointly with sociodemographic block $S$ (log population, shares aged $<30$ and $\ge 60$, and affluence).
Network intensities alone yield leave-one-out $R^2$ of $0.20$, $0.70$, and $0.61$ and sociodemographics alone $0.52$, $0.84$, and $0.54$, while the joint model reaches $0.61$, $0.94$, and $0.75$, respectively, corresponding to an additional $18\%$, $63\%$, and $43\%$ of the variance remaining after sociodemographic adjustment (see Figure \ref{fig_2}\textbf{c}).
To ensure that age composition does not confound or attenuate the network signal in age-aggregated rates, we repeat the analysis using 2024 redemption rates restricted to residents aged $\ge 60$, excluding age composition from $S$.
Here, the network alone yields $R^2$ of $0.27$, $0.58$, and $0.50$, and the joint model reaches $0.49$, $0.61$, and $0.46$, with the sociodemographic baseline explaining less than $0.07$ in all cases.
For C10, the joint model ($0.46$) falls slightly below the network alone ($0.50$), as adding uninformative features introduces noise that can penalize out-of-sample performance.
Thus, local co-purchase cohesion accounts for approximately $49\%$, $58\%$, and $48\%$ of the residual variance, respectively, with a similar residual contribution across both diabetes management and cardiovascular disease burden.
Recomputing municipal community intensities $I_c$ on the fixed-$N^{*}$ basket-size-matched samples yields nearly identical out-of-sample predictive performance ($R^2_{S+I}$ of $0.59$, $0.93$, and $0.73$ across all ages and $0.43$, $0.57$, and $0.47$ among residents aged $\ge 60$) and preserves the sign of all headline community associations (see Supplementary Materials, Section 2).

\textbf{Disentangling co-purchase cohesion from baseline product volumes.}
The lift normalizes co-occurrence by product-level frequencies, but it is still possible for $I_c$ to covary with how strongly a community is purchased across municipalities.
To account for this effect, we compare $I_c$ to both local popularity, $M=\mathrm{mean}(\log n_i+\log n_j)$, and co-purchase residualized on local product prevalence, $\widetilde{P}$, both computed on the same BiCM-validated intra-community edges.
On age-aggregated rates, with municipal log population, age composition, and affluence as controls, $I_c$ and $M$ yield similar fits at the block level ($S+I$ vs.\ $S+M$: $R^2\approx0.61$ vs.\ $0.60$ for A10; $0.94$ vs.\ $0.95$ for C; $0.75$ vs.\ $0.76$ for C10).
We therefore further examine $\widetilde{P}$, obtained by partialling $\mathrm{mean}(\log P_{ij})$ on $\mathrm{mean}(\log P_i)+\mathrm{mean}(\log P_j)$ across municipalities.
This strategy allows us to separate variation in co-purchasing from variation in local product prevalence while retaining the same BiCM-defined network structure.
After $S$, we find that $\widetilde{P}$ still adds $\Delta R^2\approx0.12$, $0.07$, and $0.11$ for A10, C, and C10 ($R^2_{S+\widetilde{P}}\approx0.64$, $0.91$, and $0.66$), closing about a quarter of the variance left unexplained by sociodemographics for A10 and C10 and nearly half for C ($26\%$, $46\%$, and $25\%$; see Figure~\ref{fig_2}\textbf{d}).
At the community level, municipalities where Candy \& snacks and Cheese \& cold cuts are more cohesively co-purchased have a higher share of residents on diabetes medication (A10), for both $I_c$ and $\widetilde{P}$ and after conditioning on sociodemographic factors and local product prevalences, while Wine \& beer cohesion tracks cardiovascular and lipid-lowering medication (C and C10) (see Supplementary Materials, Section 4 for the community-level regression coefficients).
\subsection*{Ultra-processed food prevalence and network position in co-purchasing networks}
\textbf{The share of ultra-processed food shows no urban-to-peripheral gradient.}
Because ultra-processed foods (UPFs) \cite{monteiro_ultra-processed_2019} reshape household procurement and concentrate along spatial retail gradients \cite{coyle_socio-economic_2022,scapin_global_2025}, we test whether the urban-to-peripheral rewiring of co-purchasing networks reflects differences in UPF purchase prevalence or in the topological position of ultra-processed products.
To quantify processing at the product level, we use FoodProX scores from the retail-Frida mapping \cite{menichetti_machine_2023,ravandi_prevalence_2025}, which separates NOVA 1 from NOVA 4 products with $78.3\%$ accuracy in our catalog (see Methods and Supplementary Materials, Section 5).
The UPF share of purchases varies little across municipalities (from $0.46$ to $0.56$) and is unrelated to distance from the urban core ($\rho = -0.02$, $p > 0.05$).
This holds whether products are labeled with hard NOVA classes or with FoodProX's continuous scores (class probabilities, $F\text{Score}$, high-confidence labels only, or resampled labels), and the choice does not change any of the results below.

\begin{figure*}[!htbp]
    \centering\includegraphics[width=1.0\textwidth]{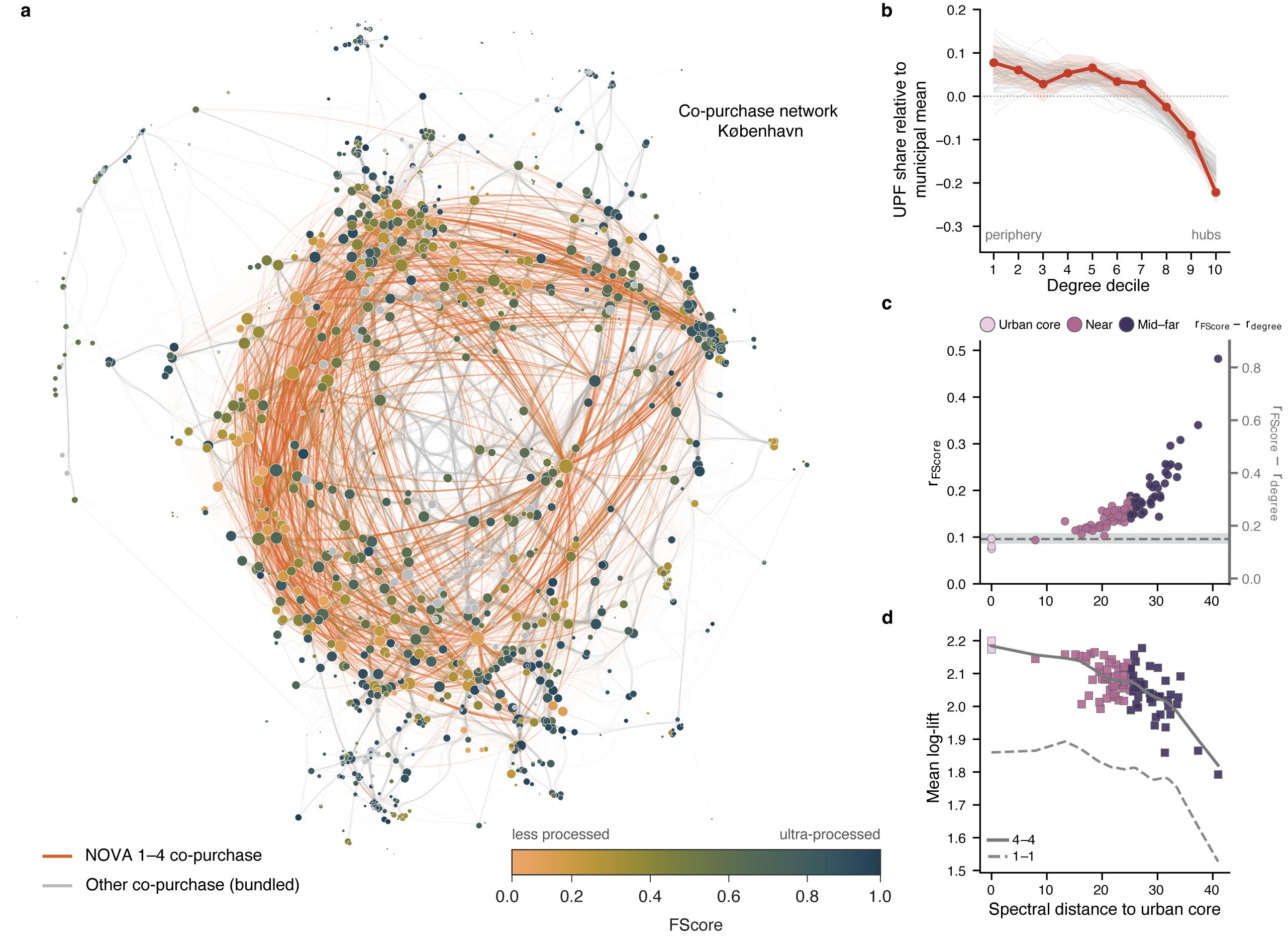}
    \caption{\textbf{Ultra-processed foods are located at the periphery of co-purchasing networks across the country.}
    \textbf{a)} Co-purchasing network for K{\o}benhavn. Node color shows the processing score $F\text{Score}$ (gray where no score is available), and node size scales with the logarithm of degree.
    Orange edges are co-purchases between minimally processed (NOVA 1) and ultra-processed (NOVA 4) products (6{,}012 links, $27\%$ of all edges), with opacity increasing with local co-purchase strength; all other links are shown in gray and edge-bundled.
    Minimally processed products form the hubs of the network, while UPFs are linked to them throughout rather than forming a separate cluster.
    \textbf{b)} UPF share within each degree decile, from the periphery (1) to the hubs (10), relative to the municipality's overall UPF share (gray lines, 95 municipalities; red, mean).
    UPFs are under-represented among the most connected products; the NOVA 4 versus NOVA 1 degree AUC is below $0.5$ in every municipality (mean $0.43$).
    \textbf{c)} Processing assortativity $r_{\mathrm{FScore}}$, the tendency of products with similar processing scores to be co-purchased, against spectral distance from the urban core.
    It rises along the urban-to-peripheral gradient ($\rho=+0.89$, $p<0.001$) and exceeds a null model that shuffles $F\text{Score}$ across products in all 95 municipalities ($z>2$).
    Its excess over degree assortativity, $r_{\mathrm{FScore}}-r_{\mathrm{degree}}$ (dashed line and band, right axis), stays positive but shows no gradient.
    \textbf{d)} Local strength (mean log-lift) of the nationally validated NOVA 4--4 pairs (points, solid line) and NOVA 1--1 pairs (dashed line). Both decline with spectral distance ($\rho=-0.56$, $p<0.001$ for 4--4; $\rho=-0.27$, $p=0.007$ for 1--1), while the 4--4 over 1--1 gap remains positive in all 95 municipalities and shows no significant gradient ($\rho=-0.16$, $p=0.13$).
    }
    \label{fig_3}
\end{figure*}

\textbf{Ultra-processed foods are located at the periphery of the co-purchasing network.}
Receipts containing both a NOVA 1 and a NOVA 4 item hold $76.6\%$ of mapped food purchase volume (see Supplementary Materials, Section 6).
Despite this within-basket co-occurrence, UPFs and minimally processed foods do not occupy the same structural role in local co-purchasing network topologies (see Figure \ref{fig_3}\textbf{a} for K{\o}benhavn).
When products are ranked into degree deciles, UPFs are under-represented among the most connected products in all 95 municipalities (see Figure \ref{fig_3}\textbf{b}).
The probability that a randomly chosen UPF has more co-purchase partners than a randomly chosen minimally processed food (the NOVA 4 versus NOVA 1 degree AUC) is below $0.5$ in every municipality (mean $\mathrm{AUC}=0.43$), indicating that UPFs consistently occupy peripheral rather than hub-like positions in local food networks (see Figure \ref{fig_3}\textbf{b}).
This peripheral role does not imply purchasing segregation across separate shopping events.
Evaluating the baskets underlying each co-purchasing tie reveals that the minimally processed (NOVA 1--1) core of the network is supported by baskets where ultra-processed items already account for approximately $35\%$ of purchase volume.
By contrast, NOVA 4--4 ties occur in baskets with nearly $60\%$ UPFs, indicating that the UPF--UPF pole concentrates within carts that are already UPF-dense rather than defining an exclusive UPF basket type (see Supplementary Materials, Section 6).
In other words, shoppers do not make separate \emph{healthy} and \emph{junk} trips, as UPFs travel in the same baskets as fresh food, but they are rarely part of the specific combinations that hold the network together, and they pair up with each other mainly in the most UPF-heavy baskets.

\textbf{Along the geographic gradient, network structure reorganizes but ultra-processed food positioning remains invariant.}
We next consider whether this shared UPF peripheral position is associated with geographic differences in how processed products connect and in how strongly UPF--UPF relationships are expressed.
Neither the co-purchase mass of NOVA 4--4 pairs ($C_{44}$) nor their excess co-prevalence above product-frequency expectations ($\Delta C_{44}$) shows a significant geographic gradient ($\rho=0.06$ and $\rho\approx 0$, respectively; $p>0.05$), indicating that aggregate UPF--UPF co-purchasing does not systematically rise or fall along the spectral axis.
In contrast, assortativity in processing scores, $r_{\mathrm{FScore}}$, increases strongly with spectral distance from the urban core ($\rho=+0.89$, $p<0.001$; see Figure \ref{fig_3}\textbf{c}), alongside rising degree assortativity.
Relative to a node-label shuffle of $F\text{Score}$ on the same graphs, $r_{\mathrm{FScore}}$ exceeds the null in all 95 municipalities ($z>2$).
The excess assortativity beyond degree, $r_{\mathrm{FScore}}-r_{\mathrm{degree}}$, remains positive across municipalities but shows no geographic gradient, indicating that the processing-specific component of assortative mixing is broadly shared nationwide (see Figure \ref{fig_3}\textbf{c}).
Finally, holding fixed the nationally validated NOVA 4--4 edges and reweighting them by local mean log-lift, association strength decreases with spectral distance ($\rho=-0.56$, $p<0.001$), while remaining higher than that of NOVA 1--1 pairs in all 95 municipalities (see Figure \ref{fig_3}\textbf{d}).
NOVA 1--1 pairs weaken as well ($\rho=-0.27$, $p=0.007$), and the gap between the two shows no gradient ($\rho=-0.16$, $p=0.13$).

Thus, across the country, UPF purchasing remains invariant in peripheral network position ($\mathrm{AUC} < 0.5$ in all 95 municipalities), aggregate UPF--UPF volume, and the higher log-lift of NOVA 4--4 over NOVA 1--1 pairs across all 95 municipalities.
Under fixed-$N^*$ sampling, processing assortativity $r_{\mathrm{FScore}}$ continues to increase with unpadded shared-product spectral distance from K{\o}benhavn ($\rho = +0.210$, $p = 0.041$; $\rho = +0.38$, $p < 0.001$ on padded spectra), while degree assortativity decreases or flattens ($\rho = -0.312$, $p = 0.002$ unpadded; $\rho = -0.11$, $p = 0.31$ padded). 
Consequently, excess processing assortativity $r_{\mathrm{FScore}} - r_{\mathrm{degree}}$ increases with distance ($\rho = +0.419$, $p < 0.001$; see Supplementary Materials, Section 2).
Repeating the analyses with Frida IDs as nodes in place of retail products recovers the same patterns, with UPF purchase share remaining uncorrelated with spectral distance across the 95 municipalities ($\rho=-0.04$, $p>0.05$); on local networks for 30 municipalities spanning the same gradient, UPFs remain peripheral (mean $\mathrm{AUC}=0.34$, all $<0.5$) while $r_{\mathrm{FScore}}$ still increases with spectral distance ($\rho=+0.89$).
\section*{Conclusion}
As research efforts increasingly focus on epidemiological and policy frameworks for ultra-processed foods, the field remains polarized between defenders of the NOVA classification \cite{monteiro_ultra-processed_2025,the_lancet_ultra-processed_2025} and critics pointing to taxonomic ambiguities and reporting artifacts in dietary recall \cite{lichtenstein_2026_2026, oneill_ultra-processed_2026, khandpur_validity_2026, louie_ultraprocessed_2026}.
Here, we address both taxonomic and observational limitations by pairing continuous processing scores \cite{menichetti_machine_2023} with passive, population-scale transaction data, leveraging 51 million grocery receipts across 95 Danish municipalities to link real-world purchasing behavior to population health outcomes \cite{bandy_use_2019,nevalainen_large-scale_2018}.
Rather than treating products in isolation, we use a Bipartite Configuration Model to isolate non-random co-purchasing patterns from popularity-driven co-occurrence \cite{saracco_randomizing_2015,saracco_inferring_2017}.
We show that while municipal UPF purchase share spans a narrow band (0.46--0.56) uncorrelated with the urban-to-peripheral spectral axis, co-purchasing networks are structurally reorganized along that gradient, and UPFs sit at the periphery of these networks everywhere.
Importantly, the structural cohesion of these local networks predicts municipal cardiometabolic prescription rates, accounting for residual variance beyond standard sociodemographic factors and product prevalence \cite{aiello_large-scale_2019}.
Together, our results indicate that geographic cardiometabolic variation reflects not only \emph{what} is purchased locally, but also \emph{how} products are combined in shopping baskets \cite{mozaffarian_history_2018,the_lancet_ultra-processed_2025}.

While our passive-data approach offers a scale that surveys cannot match, it entails inherent methodological limitations.
First, receipts record what a household buys, not what each person eats: they miss food waste, sharing within the household, and food bought elsewhere -- in other supermarket chains, restaurants or canteens \cite{nevalainen_large-scale_2018, aiello_large-scale_2019}.
Second, our design is cross-sectional and correlational, so it does not establish a causal link between co-purchasing network structure and cardiometabolic outcomes.
Third, although Danish population registries enable individual-level life-course analyses \cite{savcisens_using_2023}, we have not linked purchases and prescriptions at the individual level: because we aggregate by municipality, the households generating local baskets need not be the same residents redeeming cardiometabolic prescriptions.
Unmeasured confounders, such as physical activity, are also absent from transaction records.
Nevertheless, the reported associations remain robust to spatial cross-validation, receipt downsampling, age-restricted endpoints, Frida-ID node definitions, and filtering against maximum-entropy bipartite null models \cite{saracco_inferring_2017} (see Supplementary Materials, Sections 2--5).

Moving beyond product-centric nutrition policy requires a complex-systems view of dietary behavior and food environments \cite{rutter_need_2017, mozaffarian_history_2018}.
Current instruments, such as sugar taxes or front-of-pack warning labels, judge one product at a time \cite{mozaffarian_history_2018, the_lancet_ultra-processed_2025}.
Our results suggest that health tracks how foods are combined, not only which foods are bought -- and a rule that looks at single products cannot see those combinations.
Retailers could complement item-level rules by acting on the combinations themselves, for example through which products are placed or promoted together, or through price promotions that bundle healthier ingredients \cite{scapin_global_2025, ravandi_prevalence_2025, taillie_deal_2017}.
Future work can close the gaps listed above by following co-purchasing networks over time in longitudinal retail panels \cite{nevalainen_large-scale_2018} and by linking household purchases to individual health trajectories in Danish registry settings \cite{aiello_large-scale_2019, savcisens_using_2023} -- testing whether intervening on food combinations, not only on single products, shifts cardiometabolic risk at scale \cite{rutter_need_2017, the_lancet_ultra-processed_2025, hunter_social_2017}.
\section*{Methods}
\subsection*{Data Descriptions}
Transaction records were provided by Salling Group for Netto discount stores in Denmark from 1 February to 20 April 2025 (see Supplementary Materials, Section 1).
Each line records an encrypted basket identifier, store code, calendar day, and product metadata.
We discarded non-food lines and excluded the Easter weekend (17--20 April 2025), leaving 75 days of data.
Edible items were aggregated to brand $\times$ level-3 microcategory nodes (lowest level in the hierarchy), yielding 3,405 products after exclusion.
Store coordinates were assigned to municipal boundaries; analyses are restricted to the 95 of Denmark's 98 municipalities with complete coverage (Fan{\o}, L{\ae}s{\o}, and Bornholm excluded).
To retain population scale while standardizing sampling intensity, we downsample receipts in proportion to municipal population, anchoring the baskets-per-resident ratio at the 25th percentile of coverage; municipalities below this target retain all receipts.
We have performed sensitivity analyses varying this coverage ratio (10th--50th percentiles) and the sampling seed, as well as enforcing a fixed absolute sample size ($N^{*} = 53{,}251$) and an identical basket-size distribution across all municipalities (see Supplementary Materials, Section 2).

Cardiometabolic and sociodemographic indicators for 2024 were retrieved from Statistics Denmark \cite{statistics_denmark_statbank_2026}.
We use municipal prescription-redemption rates as treated-disease proxies \cite{aiello_large-scale_2019}: the share of residents redeeming at least one medicine in ATC classes A10 (antidiabetics; treated diabetes), C (cardiovascular system), and C10 (lipid-lowering agents, a subclass of C).
ATC A10 pools insulin (A10A) and non-insulin antidiabetics (A10B), but Statistics Denmark publishes only the pooled class at municipality level.
In the same prescription register, at national and regional resolution, 92.8\% of people who redeemed an A10 medicine in 2024 also redeemed an A10B agent (range 91.5--93.7\% across the five regions; primary sector, number of users) \cite{sundhedsdatastyrelsen_medstat_2026}.
We therefore read the municipal A10 rate as treated diabetes that is predominantly type 2.
The main analysis uses age-aggregated rates -- the number of residents who redeemed at least one relevant prescription, divided by the municipal population -- and adjusts for a sociodemographic block $S$: log total population, the population shares aged $<30$ and $\ge 60$, and an affluence index contrasting high-income ($\ge 700{,}000$~DKK) versus low-income ($<300{,}000$~DKK) households.
As a robustness check, we repeat the analysis on rates restricted to residents aged $\ge 60$ (redeemers aged $\ge 60$ divided by the population aged $\ge 60$; {\AE}r{\o} and Langeland lack cardiovascular rates in this stratum and are dropped); because this outcome is already age-stratified, $S$ then drops the age shares and keeps log total population and affluence.
Retail purchases cover February--April 2025; health and sociodemographic registries are the most recent complete calendar year (2024), immediately preceding the sales window.
\subsection*{Retail food harmonization and validation}
Each retail product was assigned a single reference entry in the DTU Frida food-composition database \cite{poulsen_dataset_2026} using exclusively its product name and store category.
We first established an automated string-matching baseline that identifies the Frida entry with highest token similarity to the retail title, prioritizing food groups compatible with the store category and searching the unconstrained catalog only when no category rule applied.
We then refined assignments through a hierarchical cascade: (i) adopting the product's EFSA FoodEx2 code \cite{european_food_safety_authority_efsa_food_2015} whenever it directly matched a Frida vocabulary entry; (ii) otherwise retrieving the nearest candidate using a pretrained multilingual text embedding model (multilingual-e5-base) restricted to category-compatible Frida food groups; and (iii) retaining the initial string match only when its token Jaccard similarity exceeded the validation median of $0.15$.
For embedding-based retrieval, candidate matches were accepted only when cosine similarity met or exceeded the median threshold of $0.825$.
Residual ambiguities related to preservation or preparation state (e.g., canned tuna in oil versus water, or salted versus raw nuts) were resolved via a compact set of deterministic, name-based domain rules (see Supplementary Materials, Section 5).
We validated the mapping against Open Food Facts NOVA classifications for a subset of products across food catalog families ($n = 2{,}944$), labeling Frida foods via FoodProX Random Forest estimates \cite{menichetti_machine_2023,ravandi_prevalence_2025}, under the NOVA framework \cite{monteiro_ultra-processed_2019}.
The complete cascade -- FoodEx2 linkage, category-gated embedding retrieval, filtered string matching, and process-state rules -- achieved $68.9\%$ overall accuracy, an ultra-processed (NOVA 4) $F_1$ score of $0.822$, and $78.3\%$ accuracy on the binary unprocessed-versus-ultra-processed contrast (NOVA 1 vs. 4).
As an independent benchmark, FoodProX \cite{menichetti_machine_2023} applied directly to product-level Open Food Facts macronutrient panels (without mapping to Frida) recovered NOVA classifications with $82.7\%$ accuracy and a UPF $F_1$ of $0.908$ on items with complete protein--fat--carbohydrate records ($n = 2{,}607$).
On items with local retailer nutrition panels, FoodProX scores derived from on-pack labels and those derived from mapped Frida profiles agreed on the categorical NOVA class for $76.8\%$ of paired items (UPF $F_1 = 0.864$; median absolute difference in continuous $F\text{Score} = 0.08$).
Route-specific accuracies, error decomposition, and the resulting catalog composition are detailed in the Supplementary Materials, Section 5.
\subsection*{Bipartite Configuration Model and Network Analysis}
The Bipartite Configuration Model constructs a maximum-entropy null ensemble conditioned on the observed degree sequences, i.e., the number of unique products purchased in each basket and, conversely, the number of baskets in which each product occurs \cite{saracco_randomizing_2015,saracco_inferring_2017,cimini_statistical_2019}.
Within this ensemble, each basket--product pair is assigned a connection probability $p_{sc} = \frac{e^{-\lambda_s-\theta_c}}{1+e^{-\lambda_s-\theta_c}}$, where ${\lambda_s}$ and ${\theta_c}$ are the Lagrange multipliers enforcing the observed degree constraints on baskets and products, respectively.
For a pair of products $a$ and $b$, the probability that both occur in basket $s$ is $q_s=p_{sa}p_{sb}$, giving an expected co-occurrence $\mu_{ab} = \langle V_{ab}\rangle = \sum_s q_s$ and variance $\sigma^2_{ab} = \sum_s q_s(1-q_s)$.
The observed co-occurrence is standardized relative to this null expectation as $z_{ab} = \frac{V_{ab}-\mu_{ab}}{\sqrt{\sigma^2_{ab}}}$.

We then assessed the statistical significance of each observed co-occurrence using the upper tail of its Poisson--Binomial null distribution \cite{saracco_inferring_2017}.
For the exact formulation, the $p$-value was computed as $p_{ab}^{\mathrm{exact}} = 1-\sum_{x=0}^{V_{ab}-1}\pi(x\mid a,b)$, where $\pi(x\mid a,b)$ is the Poisson--Binomial probability of observing $x$ common baskets under the BiCM.
We additionally computed the Refined Normal Approximation (RNA), which corrects the Gaussian tail probability using the standardized third cumulant of the Poisson--Binomial distribution \cite{lugannani_saddle_1980,hong_computing_2013}, $\gamma_{ab} = \frac{\sum_s q_s(1-q_s)(1-2q_s)}{\left[\sum_s q_s(1-q_s)\right]^{3/2}}$, yielding $p_{ab}^{\mathrm{RNA}} = \bar{\Phi}(z_{ab}) +\frac{\gamma_{ab}}{6}(z_{ab}^2-1)\phi(z_{ab})$, where $\bar{\Phi}$ and $\phi$ denote the standard normal survival function and density, respectively.
Municipal backbones are validated with exact Poisson--Binomial $p$-values, while the national backbone uses the RNA approximation.
Both exact and RNA $p$-values were subjected to Bonferroni correction for multiple hypothesis testing ($\alpha=0.001$), retaining only the links whose corrected $p$-values passed this threshold to indicate co-occurrences significantly stronger than expected from the marginal degree structure alone.

We characterized municipal network structure using the spectrum of the symmetric normalized Laplacian, which captures the organization of connectivity across the network.
Municipalities were compared by the Euclidean ($L^2$) distance between their sorted eigenvalue spectra.
When graphs differed in size, the shorter spectrum was padded with $2$, the theoretical upper bound of the normalized Laplacian.
As a robustness check, we recomputed pairwise spectral distances on the induced subgraphs over the shared product set $V_i\cap V_j$ (median $|V_i\cap V_j|=780$), with equal Laplacian order and no padding, and recovered the same urban-to-peripheral axis (Spearman $\rho=0.78$).
We perform community detection on the resulting projections using Infomap, which identifies modules by modeling information flow on the network as a random walk and optimizing a map equation \cite{rosvall_maps_2008}, and Louvain, which greedily maximizes modularity \cite{blondel_fast_2008} -- a measure of the density of edges inside communities versus edges between communities.
The two partitions were compared using the Adjusted Rand Index (ARI) and Adjusted Mutual Information (AMI), both chance-corrected.
ARI measures pairwise co-assignment from the contingency matrix of community labels, whereas AMI measures shared information between the labelings; statistical significance was assessed by permutation of labels.
Because Infomap typically yields a finer partition nested within Louvain modules, AMI remains high even when ARI is only moderate, indicating agreement on mesoscale organization.
On each municipal backbone, we quantify the tendency of products with similar characteristics to connect by measuring Newman assortativity for the continuous FoodProX $F\text{Score}$ and for degree, capturing respectively the tendency of foods with similar processing levels and similar connectivity to co-occur in the network.
We further assess whether network connectivity ranks processing extremes using the degree AUC for NOVA~4 versus NOVA~1 products, where an AUC above (below) $0.5$ indicates that NOVA~4 products systematically occupy higher-degree (lower-degree) positions than NOVA~1 products.
Finally, we assess the significance of the observed $F\text{Score}$ assortativity against a node-label shuffle null model, obtaining $z>2$ in all 95 municipalities.

For each municipality with $N$ baskets, we measure pairwise product co-occurrence using log-lift $\log \text{lift}_{ij} = \log \left(\frac{N_{ij} N}{n_i n_j}\right)$, where $N_{ij}$ is the joint basket count and $n_i, n_j$ are marginal counts.
Projecting these local co-occurrences onto the national BiCM backbone, we summarize each municipality across the Louvain communities by characterizing the local intra-community edges along three complementary dimensions.
We quantify network intensity through the mean $\log \text{lift}_{ij}$ of the observed edges, while product popularity is captured by the mean $\log n_i + \log n_j$, reflecting the baseline marginal frequency of the products involved.
Finally, to isolate co-purchasing patterns beyond product popularity, we define residual co-purchase as the municipal residual from regressing the mean $\log P_{ij}$ across community edges on the corresponding mean $\log P_i$ and $\log P_j$ across all municipalities, where $P_{ij}=N_{ij}/N$ and $P_i=n_i/N$.
To evaluate how these features predict municipal prescription rates, we fit Ridge regressions on standardized predictors (zero mean, unit variance), selecting the penalty $\alpha$ via generalized cross-validation across a 30-point log-spaced grid ($10^{-2}$ to $10^{3}$).
We measure predictive performance by leave-one-out cross-validated $R^2$ for sociodemographic controls alone ($S$), network intensity alone ($I$), and the combined model ($S+I$), repeating this for $S+M$ and $S+\widetilde{P}$.
The incremental explanatory power of the network beyond sociodemographics is quantified by the share of residual variance after $S$ is $\frac{R^2_{S+I} - R^2_S}{1 - R^2_S}$.
\section*{Declarations}
\begin{itemize}
\item Data Availability:
Raw data of purchase are not publicly available due to privacy considerations, but are available to researchers who meet the criteria for access to confidential data, sign a confidentiality agreement and agree to work under supervision in Copenhagen.
Municipal health and sociodemographic indicators were retrieved from Statistics Denmark's StatBank (\url{https://www.statbank.dk/}) via the open REST API (\url{https://api.statbank.dk/}).
Prescription-redemption counts come from table MEDI1 (\url{https://www.statbank.dk/MEDI1}; year 2024; unique persons redeeming at least one prescription in ATC classes A10, C and C10).
The regional split of antidiabetic users into insulin (A10A) and non-insulin (A10B) agents comes from Medstat (\url{https://medstat.dk/}; primary sector, number of users, 2024).
Resident population and age structure come from table BY2 (\url{https://www.statbank.dk/BY2}; 1 January 2024), and household disposable-income brackets from table INDKF132 (\url{https://www.statbank.dk/INDKF132}; year 2024).
The Danish food-composition database Frida v5.5 is publicly available from Technical University of Denmark (\url{https://frida.fooddata.dk/}).
Open Food Facts was accessed via its REST API \url{https://openfoodfacts.github.io/openfoodfacts-server/api/}.
Continuous food-processing scores $F\text{Score}$ and NOVA class probabilities were computed with FoodProX \cite{menichetti_machine_2023} (\url{https://doi.org/10.1038/s41467-023-37457-1}; code at \url{https://github.com/menicgiulia/MLFoodProcessing}), retrained on a 52-nutrient USDA panel aligned to Frida v5.5 and applied to each mapped Frida food; a seven-macronutrient FoodProX model was used for product-level Open Food Facts and on-pack information.
\item Code Availability:
The code to reproduce the analysis is released on \href{https://github.com/RiegelGestr/FoodHealth}{FoodHealth}.
For inquiries, please contact A.D. \url{antde@dtu.dk}.
\item Acknowledgements
A.D. acknowledges helpful discussions and insightful feedback from Riccardo Di Clemente, Sándor Juhász, Anna Mancini, and Fabio Saracco.
All authors acknowledge helpful feedback from Ulf Aslak at an early stage of this work.
Finally, all authors thank the BI \& Data Analytics team (Machine Learning \& AI) at Salling Group, with special thanks to Philip Lund Thomsen for initiating and driving this collaboration.
\item Author Contributions:
L.R. performed preliminary analysis.
A.D. performed the analysis and made the figures.
A.D. and S.L. designed the analysis.
S.L. supervised the project.
All authors discussed the results and contributed to the final manuscript.
\item Competing Interests:
The authors declare no competing interests.
\item Funding Declaration:
S.L. acknowledges support from the Villum Foundation through the Nation-Scale Social Networks project (grant number 34288).
A.D. acknowledges support by COLINE DUT European Partnership project (F-DUT-2023-0037) and the Lagrange Project of the ISI Foundation, funded by Fondazione CRT.
\end{itemize}
\bibliography{bibliography}
\pagebreak[2]
\clearpage
\renewcommand{\figurename}{Supplementary Figure}
\renewcommand{\tablename}{Supplementary Table}
\setcounter{figure}{0}
\renewcommand{\thefigure}{S\arabic{figure}}
\renewcommand{\theHfigure}{S\arabic{figure}}
\section*{Supplementary Materials, Section 1: Dataset Information}
\begin{table}[h!]
	\centering
	\begin{tabular}{lp{0.62\textwidth}}
		\toprule
		{\textbf{Column}} & {\textbf{Description}}\\
		\midrule
		Store ID & Encrypted identifier uniquely identifying each store (\texttt{store\_id\_encrypted}). 553 stores are assigned to a municipality. \\
		Latitude & Geographic latitude of the store (WGS84), taken from the checkout logs. \\
		Longitude & Geographic longitude of the store (WGS84), taken from the checkout logs. \\
		\hline
	\end{tabular}
	\caption{Store metadata used to locate checkouts in space and to aggregate baskets by municipality.}
	\label{tab_sm_data_store}
\end{table}
\begin{table}[h!]
	\centering
	\begin{tabular}{lp{0.62\textwidth}}
		\toprule
		{\textbf{Column}} & {\textbf{Description}}\\
		\midrule
		Municipality & Official municipality name, aligned to StatBank nomenclature. \\
		Population & Residential population used for per-capita downsampling of baskets. \\
		Share under 30 & Share of residents aged 0--29 years (BY2). \\
		Share over 60 & Share of residents aged 60 years and older (BY2). \\
		Affluence index & $(N_{\ge 700\mathrm{k}}-N_{<300\mathrm{k}})/N_{\mathrm{families}}$ from disposable family income (INDKF132). \\
		Rate A10 & Unique redeemers of antidiabetics (ATC A10) per resident (MEDI1, 2024). \\
		Rate C & Unique redeemers of cardiovascular drugs (ATC C) per resident (MEDI1, 2024). \\
		Rate C10 & Unique redeemers of lipid-modifying agents (ATC C10) per resident (MEDI1, 2024). \\
		Age stratum & For age-specific rates: 0--17, 18--29, 30--59, and $\ge 60$ years. \\
		\hline
	\end{tabular}
	\caption{2024 StatBank sociodemography and prescription redemption rates for each municipality.
	Age-aggregated rates use the municipal population as denominator; analyses among residents $\ge 60$ years use the corresponding age-specific rates.}
	\label{tab_sm_data_kommune}
\end{table}
\begin{table}[h!]
	\centering
	\begin{tabular}{lp{0.62\textwidth}}
		\toprule
		{\textbf{Column}} & {\textbf{Description}}\\
		\midrule
		Receipt ID & Encrypted basket identifier. One ID is one checkout. \\
		Product ID & Encrypted Product ID. \\
		Store ID & Encrypted store at which the basket was recorded. \\
		Date & Calendar day of the checkout. \\
		\hline
	\end{tabular}
	\caption{Bipartite basket--product incidence.
	Each row is a purchased item.
	The bipartite graph has baskets on one side and products on the other; an edge exists if the product appears in the checkout.}
	\label{tab_sm_data_bipartite}
\end{table}
\begin{table}[h!]
	\centering
	\begin{tabular}{lp{0.62\textwidth}}
		\toprule
		{\textbf{Column}} & {\textbf{Description}}\\
		\midrule
		Product ID & Encrypted product ID. \\
		Crosswalk status & Mapped, Unmapped, or Non-Food. Nutrient joins use Mapped only. \\
		Frida food ID & DTU Frida~5.5 FoodID. \\
		Food group & Frida food group (DTU taxonomy, distinct from the retailer tree). \\
		Energy & Energy (kcal per 100~g). \\
		Macronutrients & Protein, fat, available carbohydrate, and dietary fiber (g per 100~g). \\
		Sodium & Sodium (mg per 100~g). \\
		Added / free sugar & Added sugar and free sugars (g per 100~g). \\
		Saturated fat & Sum of saturated fatty acids (g per 100~g). \\
		NOVA (FoodProX) & FoodProX NOVA group $\{1,2,3,4\}$. \\
		FScore & FoodProX continuous processing score (higher $=$ more processed). \\
		\hline
	\end{tabular}
	\caption{Product--nutrition link: hybrid retail-to-Frida crosswalk and Frida~5.5 fields used in the analyses.}
	\label{tab_sm_data_frida}
\end{table}
\begin{table}[h!]
	\centering
	\begin{tabular}{lp{0.62\textwidth}}
		\toprule
		{\textbf{Macro-category}} & {\textbf{Meso-categories}}\\
		\midrule
		Fresh food &
		Chilled ready meals \& deli;
		dairy, milk \& eggs;
		meat, pork \& poultry;
		fresh bakery \& bread;
		fruits;
		vegetables \& greens;
		seafood \& fish \\
		Frozen foods &
		Ice cream \& frozen desserts;
		frozen meals \& pizza;
		frozen meat \& seafood;
		frozen vegetables \& fruits \\
		Pantry \& packaged food &
		Baking supplies \& flour;
		sauces, dips \& spreads;
		oils, vinegars \& condiments;
		canned goods \& soups;
		cereals, oats \& breakfast;
		pasta, rice \& grains \\
		Snacks \& sweets &
		Cookies, cakes \& sweet biscuits;
		nuts, seeds \& dried fruit;
		energy bars \& snacks;
		chips, crisps \& salty snacks;
		chocolate, candy \& sweets \\
		Beverages &
		Wine, spirits \& alcohol;
		juices, smoothies \& soft drinks;
		beer \& cider;
		coffee, tea \& hot drinks;
		water (still \& sparkling) \\
		\hline
	\end{tabular}
	\caption{Food meso-categories used in the analyses, grouped by macro-category.
	This is the mapped taxonomy by the data provider.}
	\label{tab_sm_data_cat_map}
\end{table}
\pagebreak[2]
\clearpage
\section*{Supplementary Materials, Section 2: Downsampling stability}
To assess the stability of the geographic pattern, we repeated the population-scaled downsampling while varying the basket-per-person ratio and the random seed used to sample receipts.
Both analyses used the RNA approximation to the Bipartite Configuration Model with Bonferroni correction at $\alpha=0.001$, matching the procedure used in the main analysis \cite{saracco_randomizing_2015,saracco_inferring_2017,cimini_statistical_2019}.

\begin{figure}[h!]
\centering
\includegraphics[width=\textwidth]{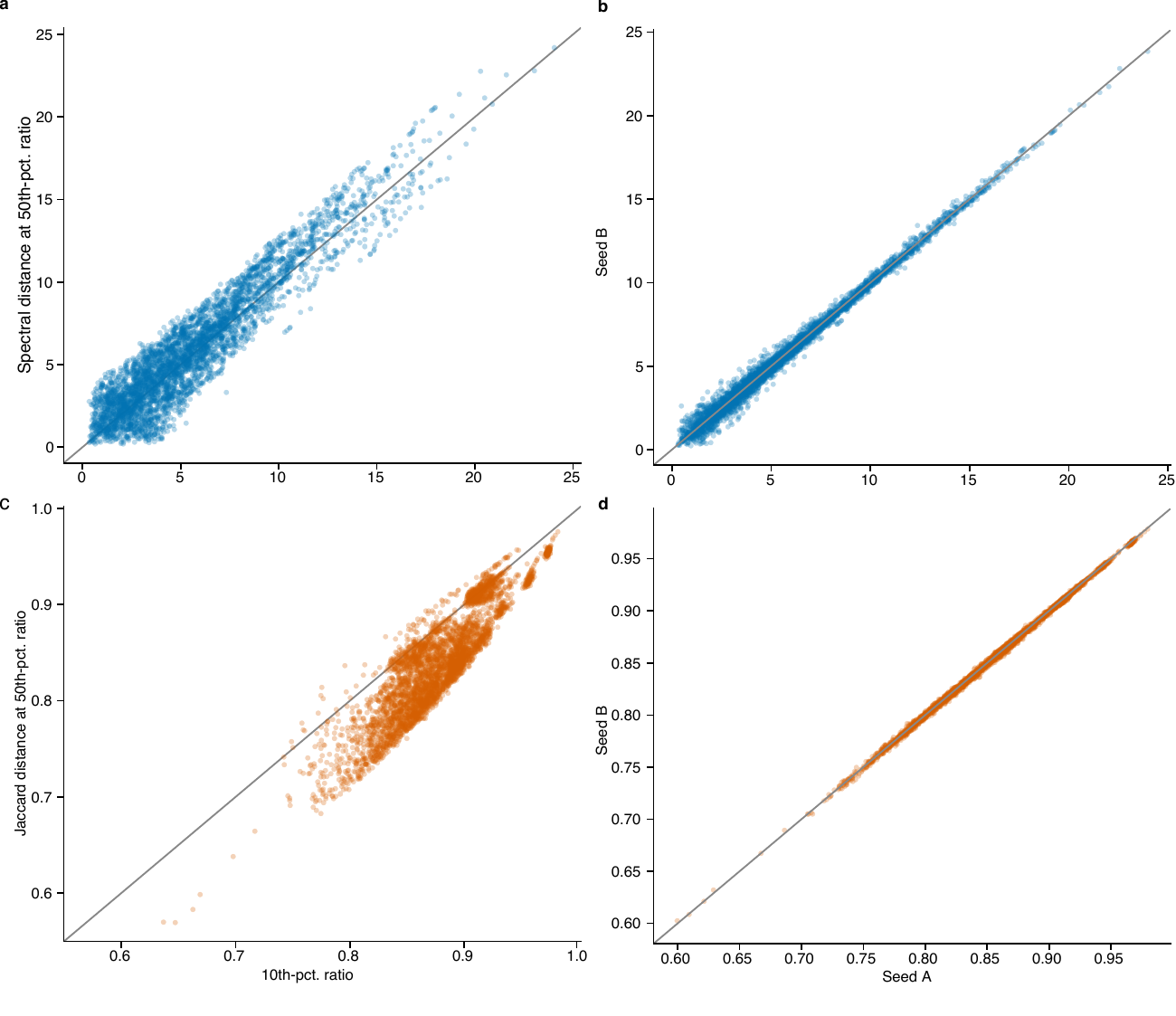}
\caption{\textbf{Downsampling stability of municipal spectral and Jaccard geography.}
Pairwise distances among 95 municipalities (upper triangle of the distance matrix) after population-scaled RNA backbones (Bonferroni $\alpha=0.001$).
In all panels, the gray line is the identity $y=x$.
\textbf{a)} Spectral distance (normalized Laplacian) at the 10th versus 50th percentile of municipal basket/person coverage (Spearman $\rho=0.907$).
\textbf{b)} Spectral distance for two independent draws at the 25th-percentile ratio (Spearman $\rho=0.993$).
\textbf{c)} Pairwise edge Jaccard distance at the same 10th versus 50th coverage percentiles (Spearman $\rho=0.878$).
\textbf{d)} Pairwise edge Jaccard distance for the same two independent seeds (Spearman $\rho=0.999$).
The urban--peripheral ranking is recovered under both perturbations.
}
\label{sm_downsample_stability}
\end{figure}

Within each municipality, a single receipt permutation was used to construct nested samples truncated at the 10th, 25th, and 50th percentiles of municipal coverage, corresponding to $\approx3.80$, $5.41$, and $7.73$, respectively.
The resulting pairwise spectral-distance matrices were highly correlated, with Spearman $\rho=0.973$ between the 10th and 25th percentiles, $\rho=0.965$ between the 25th and 50th percentiles, and $\rho=0.907$ between the 10th and 50th percentiles (see Figure \ref{sm_downsample_stability} \textbf{a}).
The same pattern was observed for the urban-to-peripheral axis, defined as the mean spectral distance of each of the remaining 91 municipalities from the four-city core. The rankings obtained at the three basket-density levels had Spearman correlations of $\rho=0.989$, $0.982$, and $0.955$ for the same three pairwise comparisons.
A median split of this axis, corresponding to the near/mid--far classification used in the main text, gave adjusted Rand indices of $0.83$ between adjacent ratios and $0.68$ between the 10th and 50th percentiles.

We next evaluated the effect of the random seed used for receipt sampling.
An analysis of 10 municipalities, including the four cities together with rural municipalities, was repeated for 100 independent draws at the production 25th-percentile ratio.
Across the $4{,}950$ pairs of runs, the pairwise spectral-distance matrices had a mean Spearman correlation of $\rho=0.997$ (median $0.997$, range $0.991$--$1.000$).
Five independent draws on the full set of 95 municipalities gave similarly high correlations, with $\rho=0.992$--$0.996$ (see Figure \ref{sm_downsample_stability} \textbf{b}).

The corresponding urban-to-peripheral spectral axis was also highly consistent across seeds ($\rho\geq0.996$), and the near/mid--far median split was almost identical across all replicates (ARI $\sim1$).
The same stability is observed for pairwise edge Jaccard similarity (see Figure \ref{sm_downsample_stability} \textbf{c} and \textbf{d}), providing a complementary measure of the consistency of the underlying network structure.
Overall, the geographic partition is consistent across the tested downsample seeds and remains similar when the basket-density ratio is varied over approximately a twofold range, with both spectral distance and edge Jaccard similarity showing consistent network structure across downsampling choices.

The processing-assortativity gradient is computed on the municipal graphs from the main analysis, obtained using the exact BiCM with Bonferroni correction ($\alpha=0.001$), rather than on the RNA backbones used in the ratio/seed sensitivity analysis above.
Reconstructing the exact backbones at the 25th and 50th coverage percentiles gives nearly identical municipal rankings with $r_{\mathrm{FScore}}$ \cite{menichetti_machine_2023} at q25 correlating with the main-analysis values at $\rho=+0.98$, while q25 and q50 correlate at $\rho=+0.96$.
The q25 networks also retain the same ordering along the spectral axis of the main analysis ($\rho=+0.88$), with an edge-set Jaccard similarity of 0.79 relative to the main-analysis graphs.
In addition, the main-analysis $r_{\mathrm{FScore}}$ remains correlated with the RNA spectral axis reported in this Supplementary Materials, Section 2 ($\rho=+0.79$).
These results show that the urban–peripheral gradient in processing assortativity is robust to changes in basket density within the exact-BiCM construction and is not specific to the RNA sensitivity networks.

\begin{figure}[h!]
\centering
\includegraphics[width=\textwidth]{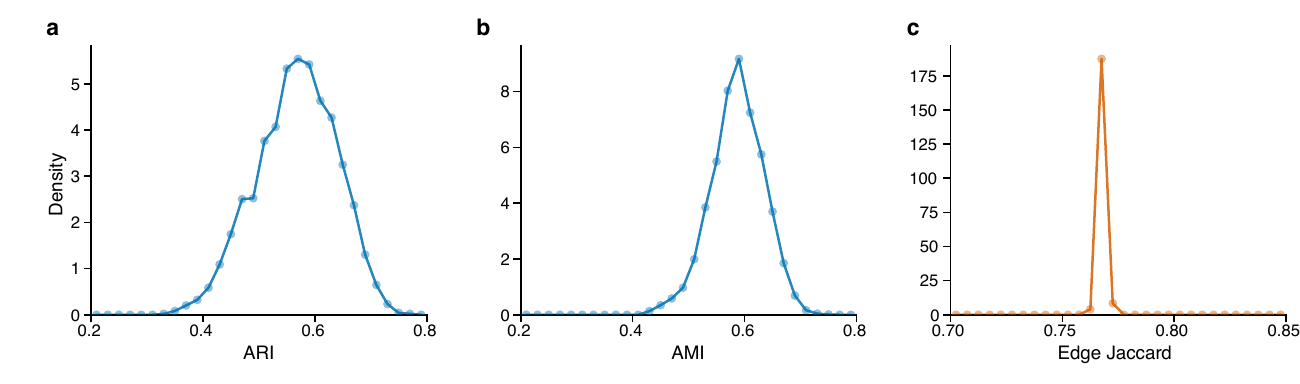}
\caption{\textbf{Stability of pooled global network backbones and community structure across downsampling seeds.}
Distributions of partition concordance and edge-set overlap across all 4,950 unordered pairs of 100 independent receipt draws on the ten-municipality pilot at the 25th-percentile basket coverage (baskets-per-resident $\approx 5.41$, RNA BiCM with Bonferroni $\alpha=0.001$).
\textbf{a)} Distribution of the Adjusted Rand Index (ARI) between Louvain community partitions (mean $0.566$).
\textbf{b)} Distribution of the Adjusted Mutual Information (AMI) for the same community partitions (mean $0.587$).
\textbf{c)} Pairwise Jaccard similarity distribution of the validated edge sets ($0.768 \pm 0.001$, 5th--95th percentile $0.766$--$0.770$). Variation in ARI and AMI reflects soft boundaries in Louvain modularity optimization rather than substantive changes in the validated edge catalog.
}
\label{sm_global_downsample}
\end{figure}

We repeated the pooled global network pipeline on the ten-municipality pilot used for local downsample checks (four urban centers plus rural municipalities).
For each of 100 independent receipt draws at the production 25th-percentile basket density (baskets-per-resident $\approx 5.41$), we pooled downsampled baskets, fitted a global RNA BiCM with Bonferroni correction at $\alpha=0.001$, and retained the validated edge catalog ($\sim$87\,000 edges per replicate).
For every unordered pair of replicates we computed the Jaccard overlap of global edge sets and the agreement between Louvain partitions on the intersection of nodes \cite{blondel_fast_2008}(Adjusted Rand Index and Adjusted Mutual Information).
Across 4,950 pairs the global edge catalog was highly reproducible (Jaccard $0.768\pm 0.001$; 5th--95th percentile $0.766$--$0.770$), and community structure was consistently preserved (mean ARI $0.566$, mean AMI $0.587$; see Figure \ref{sm_global_downsample}).
The cardiometabolic analysis is similarly robust to the downsampling seed, as repeating the main-text construction on two independent national redraws at the same 25th-percentile basket density yields comparable leave-one-out fits.
Louvain detected $\sim$7 communities per replicate (mean $7.3$, s.d.\ $0.7$).
Thus the downsample seed does not materially alter the global backbone; remaining variation in ARI/AMI reflects soft boundaries in the Louvain partition rather than wholesale restructuring of the catalog.
We observe the same pattern on the full national pool.
Comparing the main-text global network with two independent redraws at the same 25th-percentile ratio yields pairwise Jaccard $0.740$ in all three pairs, and Louvain agreement at least as high as in the previous ten-municipality pilot (mean ARI $0.67$, mean AMI $0.64$).

The global network reported in the main text uses the RNA approximation to estimate the upper-tail probabilities of the Poisson–Binomial distributions arising from the BiCM.
As a robustness check, we refitted the national pooled network using the Lugannani–Rice saddlepoint approximation \cite{lugannani_saddle_1980}, which estimates the tail probability from the cumulant generating function, evaluating both approximations directly on the full Bonferroni-validated graph ($\alpha=0.001$, prior to the $z^{*}=28$ percolation cutoff; $3{,}171$ nodes and $60{,}126$ edges).
On this unfiltered backbone, the two approximations produced nearly identical leave-one-out cardiometabolic fits ($R^{2}_{S+I}=0.63$, $0.93$, and $0.75$ under RNA versus $0.61$, $0.94$, and $0.76$ under Lugannani–Rice for A10, C, and C10, respectively, closely matching the $0.61$, $0.94$, and $0.75$ obtained at $z^{*}=28$) and an essentially identical municipal ranking of mean NOVA 4--4 log-lift (Spearman $\rho=0.998$).

In the main analysis, each municipality is sampled in proportion to its population, so larger municipalities enter the projection with more baskets $N$ and greater statistical power under Bonferroni validation.
In a direct comparison between K{\o}benhavn and Norddjurs, equalizing sample size ($N \approx 190{,}000$ baskets), applying an effect-size filter ($z/\sqrt{N}$), or permuting municipality labels across baskets within matched basket-size bins ($p < 0.01$) keeps their spectral distance well above resampling noise, whereas K{\o}benhavn remains indistinguishable from its immediate urban neighbor Frederiksberg or noticeably closer to urban centers such as Odense and Aarhus.
To apply the same control across all 95 municipalities, we fixed both the number of baskets and the basket-size distribution (number of distinct products per receipt, binned from $1$ to $39$ and $\ge 40$), setting each bin to the minimum count available across municipalities to account for the fact that smaller baskets produce fewer co-purchases and sparser backbones even at equal $N$, yielding $N^{*} = 53{,}251$ baskets per municipality.
We drew one hundred independent samples without replacement per municipality, projected each with the exact BiCM (Bonferroni $\alpha = 0.001$), and ranked municipalities by their mean spectral distance to K{\o}benhavn.

\begin{figure}[h!]
  \centering
  \includegraphics[width=\textwidth]{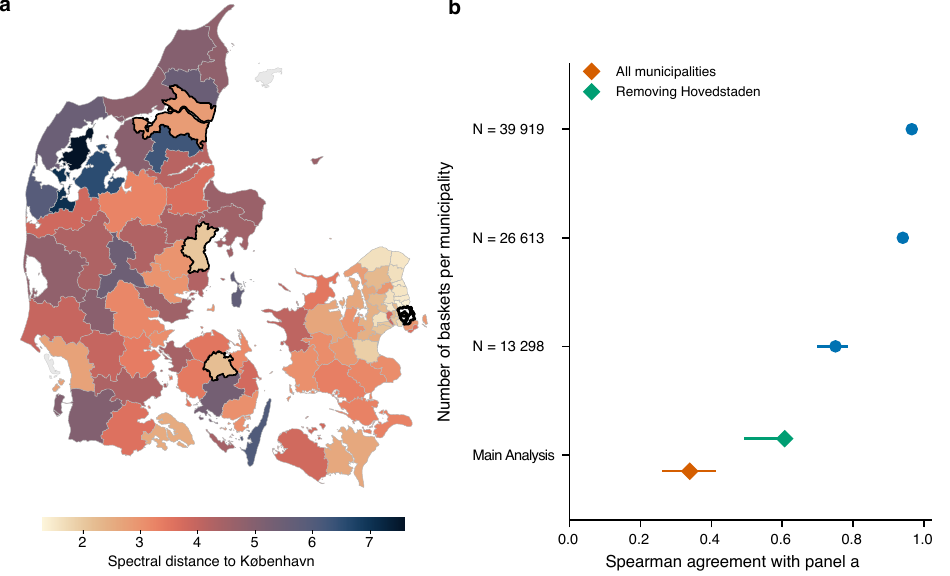}
  \caption{\textbf{Spectral distance to K{\o}benhavn at a fixed number of baskets and a fixed basket-size distribution.}
  \textbf{a)} Mean spectral distance between each municipality's backbone and K{\o}benhavn's across independent draws at $N^{*} = 53{,}251$ baskets with identical basket-size distributions (exact BiCM, Bonferroni $\alpha = 0.001$). Lighter shades are closer to K{\o}benhavn (white); Aarhus, Odense, and Aalborg are outlined; municipalities without data are gray and Bornholm is omitted.
  \textbf{b)} Spearman rank correlation with \textbf{a}.
  Blue circles indicate the same basket-size mix drawn at smaller common \(N\), with points showing the mean across draws and bars the leave-one-draw-out range. Diamonds indicate agreement with the population-scaled ranking in the main text, both across all municipalities and outside the Capital Region (Region Hovedstaden).
  }
  \label{sm_kbh_fixed_cap}
\end{figure}

Fixing $N^{*}$ and matching basket sizes brings 21 of the 27 Capital Region suburbs into the closest tertile to K{\o}benhavn, reflecting a metropolitan organization consistent with \cite{desiderio_mapping_2026} (see Figure \ref{sm_kbh_fixed_cap}\textbf{a}).
Because of this high similarity across the Capital Region, the other major urban centers (Aarhus, Odense, and Aalborg) -- although still close to K{\o}benhavn and distinct from their surrounding rural municipalities -- do not form a cleanly separated four-city cluster, which is why we measure spectral distance directly relative to K{\o}benhavn.
The resulting ranking is stable under smaller common samples, showing Spearman correlations with the $N^{*}$ ordering of $\rho = 0.97$, $0.94$, and $0.75$ for $3N^{*}/4$, $N^{*}/2$, and $N^{*}/4$ baskets, and of $\rho = 0.99$ with the effect-size filter $z/\sqrt{N} \ge 0.05$, while agreeing with the main-text ranking at $\rho = 0.61$ outside the Capital Region ($\rho = 0.34$ overall; see Figure \ref{sm_kbh_fixed_cap}\textbf{b}) and preserving a monotonic decline in edge overlap with K{\o}benhavn across fixed-$N^{*}$ tertiles ($0.44$, $0.40$, and $0.36$; $\rho = -0.85$).
Recomputing community intensities $I_c$ on these fixed-$N^{*}$ draws also recovers the main-text cardiometabolic fits ($R^{2}_{S+I}$ of $0.59$, $0.93$, and $0.73$ across all ages and $0.43$, $0.57$, and $0.47$ for residents aged $\ge 60$, with all 16 headline community coefficients keeping their sign and 13 of 16 remaining significant at $p < 0.05$).
Finally, on these fixed-$N^{*}$ backbones, UPFs remain peripheral ($\mathrm{AUC} < 0.5$, mean $0.37$) and maintain higher 4--4 than 1--1 log-lift in all 95 municipalities, while processing assortativity $r_{\mathrm{FScore}}$ increases with spectral distance from K{\o}benhavn both on padded spectra ($\rho = +0.38$, $p < 0.001$) and on unpadded shared-product spectra ($\rho = +0.210$, $p = 0.041$), whereas degree assortativity $r_{\mathrm{degree}}$ remains flat on padded spectra ($\rho = -0.11$, $p = 0.31$) and decreases on shared-product spectra ($\rho = -0.312$, $p = 0.002$), resulting in a positive gradient for excess processing assortativity $r_{\mathrm{FScore}} - r_{\mathrm{degree}}$ ($\rho = +0.419$, $p < 0.001$).
\pagebreak[2]
\clearpage
\section*{Supplementary Materials, Section 3: National co-purchase network and communities}
\begin{figure}[h!]
\centering
\includegraphics[width=\textwidth]{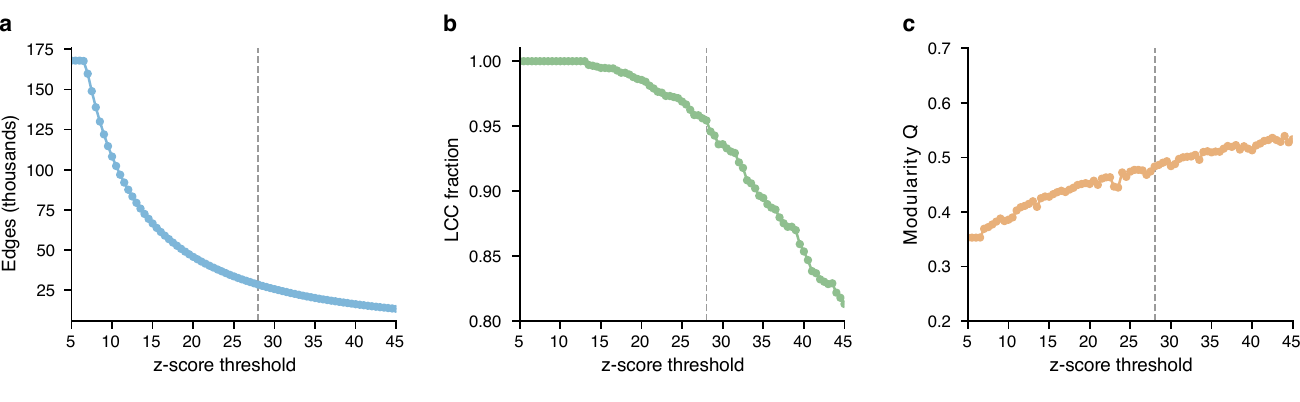}
    \caption{\textbf{Network topological properties across statistical validation thresholds.}
    \textbf{a)} Number of retained edges (in thousands) as a function of the $z$-score cutoff.
    \textbf{b)} Fraction of non-isolate nodes belonging to the largest connected component (LCC).
    \textbf{c)} Modularity $Q$ resulting from Louvain community detection ($\gamma = 1$).
    In all panels, the vertical dashed gray line marks the selected operating threshold $z^* = 28$, which balances the removal of weak cross-category ties against the onset of topological fragmentation.
}
\label{sm_zscore_sweep}
\end{figure}

To construct the national food co-purchase network, we selected a robust threshold for downstream community detection by analyzing the network topology as a function of the BiCM-derived $z$-score cutoff \cite{saracco_inferring_2017,saracco_randomizing_2015}.
As shown in Figure \ref{sm_zscore_sweep}, we tracked three key metrics: the number of preserved edges, the fraction of non-isolated nodes in the largest connected component (LCC), and the Louvain modularity $Q$ ($\gamma = 1$).

Increasing the significance threshold leads to a monotonic, power-law-like pruning of connections, reducing the edge count from over $168{,}000$ at $z=5$ down to $28{,}435$ at $z^*=28$ (an overall reduction exceeding $80\%$).
Importantly, the modularity profile (see Figure \ref{sm_zscore_sweep} \textbf{c}) reveals that the network already possesses an intrinsically modular organization at permissive thresholds ($Q \approx 0.4$ at $z=10$), indicating that product clustering reflects genuine consumption patterns rather than an artifact of the filtering \cite{blondel_fast_2008,rosvall_maps_2008}.
As $z$ increases, modularity grows steadily up to $Q \approx 0.49$ at $z^*=28$ as spurious inter-cluster ties are systematically eliminated.
In contrast, the integrity of the giant component displays a pronounced transition (see Figure \ref{sm_zscore_sweep} \textbf{b}).
While the LCC remains fully intact ($100\%$ of non-isolate nodes) up to $z \approx 13$ and degrades smoothly up to $z \approx 28$ before higher thresholds induce steep percolation-like fragmentation (dropping to $81\%$ at $z = 45$), we identified $z^* = 28$ as the optimal trade-off -- the most stringent threshold preserving $\ge 95\%$ of nodes within a single coherent component.

\begin{figure}[h!]
\centering
\includegraphics[width=\textwidth]{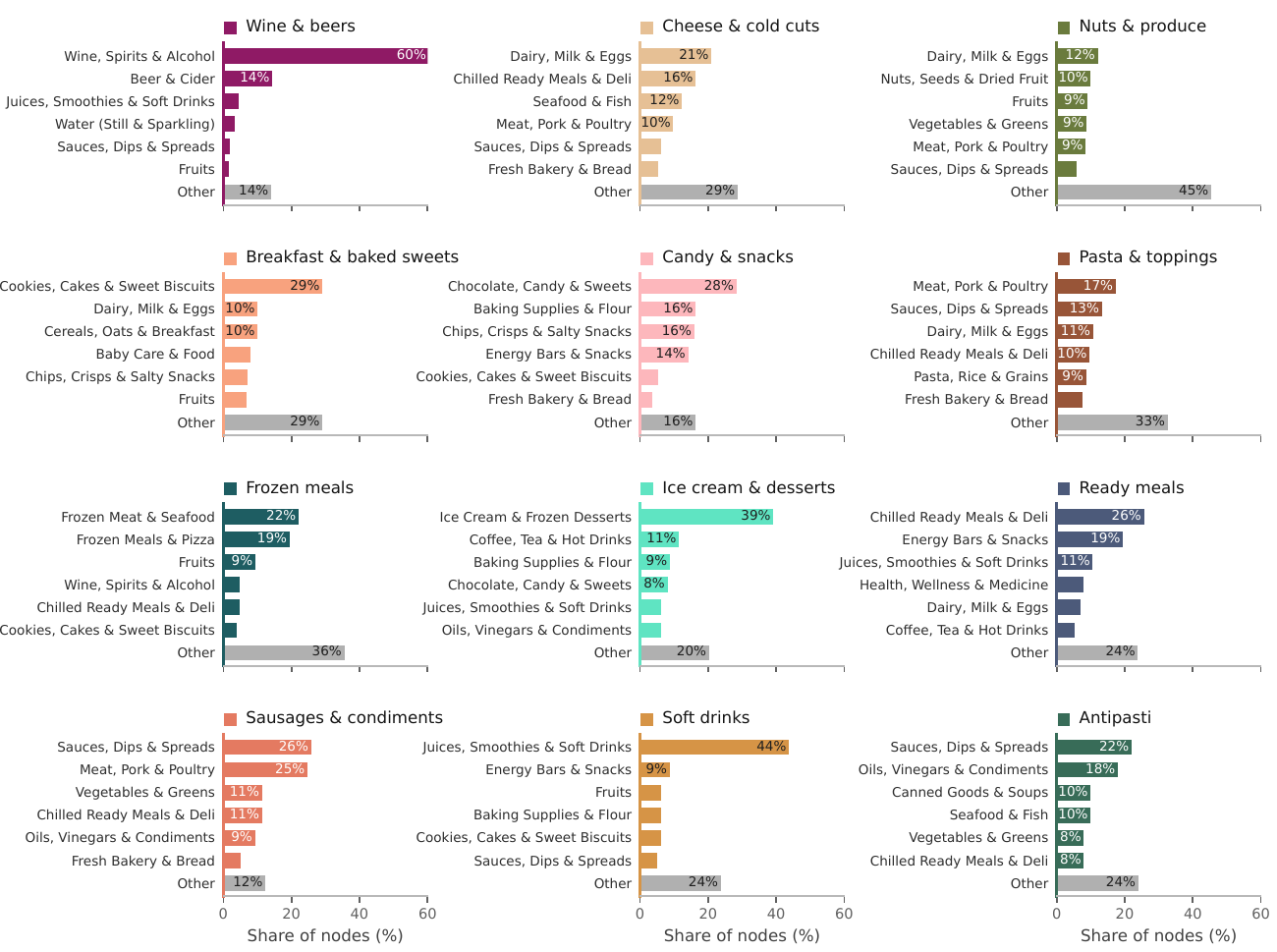}
\caption{\textbf{Product category composition of national dietary co-purchasing communities.}
Breakdown of food categories across the 12 major dietary communities identified via Louvain modularity optimization on the validated national co-purchasing backbone ($z^* = 28.0$).
Bars represent the percentage share of product nodes belonging to each food category within each community.
Minor categories are aggregated under ``Other'' and displayed in gray.
}
\label{sm_community_composition}
\end{figure}

We next characterized the composition of these communities according to the product categories of their nodes (see Figure \ref{sm_community_composition}).
For each community, we report the share of nodes belonging to its six most frequent meso-categories, while all remaining categories are grouped as Other.
The resulting partition provides the community structure used to compute municipal community intensities $I_c$ in the cardiometabolic analysis.

\begin{figure}[h!]
\centering
\includegraphics[width=\textwidth]{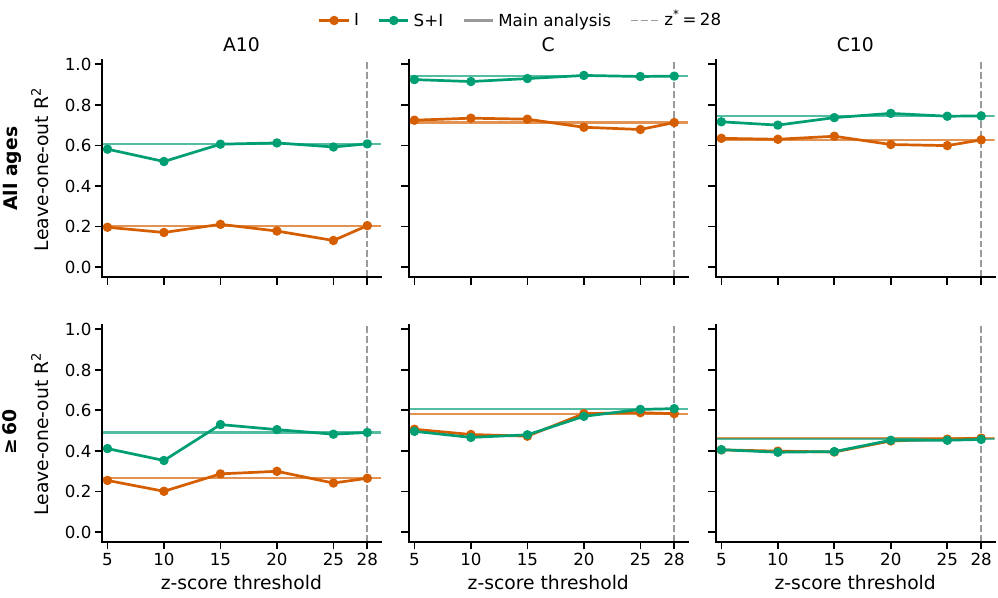}
\caption{\textbf{Cardiometabolic $I_c$ fits across $z$-score thresholds.}
Leave-one-out Ridge $R^2$ for community intensities $I$ (orange) and the joint model $S{+}I$ (green) across $z$-score thresholds, with communities re-derived at each cutoff using Louvain on the LCC.
Results are shown for age-aggregated rates (top) and rates among residents $\geq 60$ years (bottom) across A10, C, and C10.
Horizontal lines indicate the main-analysis $R^2$ at $z^{*}=28$, while the vertical dashed line marks $z^{*}=28$.
}
\label{sm_zscore_Ic_health}
\end{figure}

To assess whether the cardiometabolic association is tied to the cutoff, we repeated the $I_c$ construction at different $z$-score thresholds, recomputing Louvain communities on the LCC at each threshold rather than imposing the $z^{*}$ partition (see Figure \ref{sm_zscore_Ic_health}).
Leave-one-out $R^2$ for $I$ and $S{+}I$ recovers the main-analysis values for A10, C and C10 in both the age-aggregated and $\geq 60$ specifications, indicating that the result at $z^{*}=28$ is robust to the choice of threshold.
Repeating that evaluation on the LCC at $z\in\{10,15,20,25,28\}$, 4--4 mean log-lift exceeded 1--1 in all 95 municipalities at every threshold, and the Spearman association with spectral distance remained negative and significant for $z\geq 15$ ($\rho$ from $-0.34$ to $-0.58$, $p<0.001$), as in the main text.
\pagebreak[2]
\clearpage
\section*{Supplementary Materials, Section 4: Community-level regression coefficients and spatial robustness}
To identify the dietary communities contributing to the block-level association between local co-purchasing structure and cardiometabolic outcomes, we fitted one-community-at-a-time OLS models of municipal prescription rates on the sociodemographic block $S$ and a single community-level co-purchasing feature.
Predictors were standardized, and heteroskedasticity-robust HC3 standard errors were used throughout.
For each community, we considered either the mean intra-community log-lift $I_c$ or the residualized co-purchase measure $\widetilde{P}$.
We considered nine of the twelve major Louvain communities identified in the national backbone \cite{blondel_fast_2008}; frozen meals, soft drinks, and antipasti were excluded because fewer than 50 local intra-community edges were available for these categories in certain municipalities.

\begin{figure}[h!]
\centering
\includegraphics[width=\textwidth]{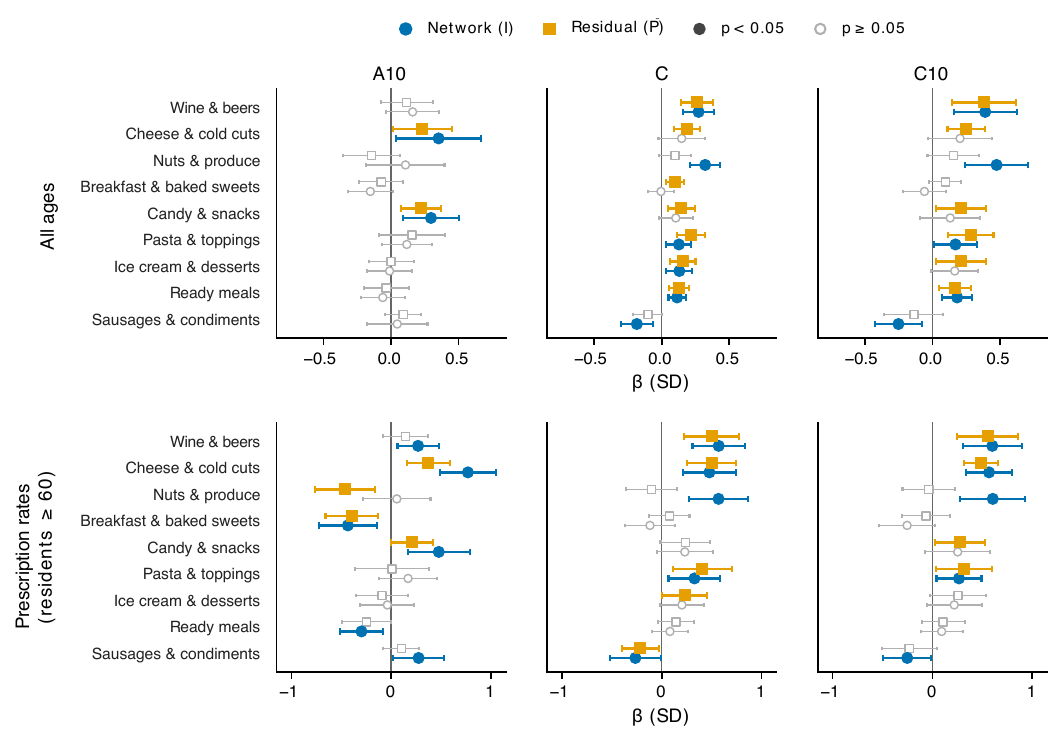}
\caption{\textbf{Community-level associations between food co-purchasing structure and cardiometabolic prescription rates.}
    Standardized OLS regression coefficients ($\beta$, in outcome standard deviations) from municipality-level models including the sociodemographic block $S$ and one community-level co-purchasing measure at a time.
    From left to right, panels show associations with antidiabetic (A10), cardiovascular (C), and lipid-modifying (C10) prescription rates, for age-aggregated rates (top) and residents aged $\geq60$ (bottom).
    Blue points show mean intra-community log-lift ($I_c$), while orange squares show residualized co-purchase ($\widetilde{P}$); horizontal bars indicate HC3 robust 95\% confidence intervals.
    Filled symbols indicate $p<0.05$ while open symbols $p\geq0.05$.
    }
\label{sm_coeff_comm}
\end{figure}

The resulting standardized regression coefficients are shown in Figure \ref{sm_coeff_comm}.
For age-aggregated rates, the clearest associations are observed for A10.
Candy \& snacks and Cheese \& cold cuts show positive associations with antidiabetic prescription rates for both measures of co-purchasing cohesion \cite{lane_ultra-processed_2024,micha_red_2010}, with coefficients of approximately $\beta=0.30$ and $0.35$ for $I_c$, and $\beta=0.22$ and $0.23$ for $\widetilde{P}$, respectively.
Wine \& beers is positively associated with both cardiovascular (C) and lipid-modifying (C10) rates, with $I_c$ coefficients of approximately $\beta=0.27$ and $0.39$, respectively, and the associations remain positive for $\widetilde{P}$.
Nuts \& produce shows comparatively large positive $I_c$ coefficients for C and C10, but these associations are not retained for $\widetilde{P}$, indicating that their apparent relationship with cardiometabolic rates is largely accounted for by local product prevalence.

The age-specific analysis shows a broadly similar pattern despite the reduced sociodemographic adjustment.
In particular, Candy \& snacks and Cheese \& cold cuts remain positively associated with A10, while Wine \& beers remains positively associated with C and C10.
Several coefficients are larger in magnitude than in the age-aggregated analysis, consistent with the substantially lower explanatory power of the sociodemographic baseline in this age group.
Overall, the community-level analysis therefore identifies a small set of dietary communities whose co-purchasing cohesion is consistently associated with cardiometabolic outcomes, while also distinguishing associations that persist after accounting for local product prevalence from those that do not.
Because the HC3 standard errors used in the regression models of the main analysis address heteroskedasticity but not residual spatial autocorrelation, we assessed Moran's $I$ on OLS residuals using row-standardized 5-nearest-neighbor weights on municipal centroids and 999 permutations, retaining the same complete-case samples as in the main analysis.
For age-aggregated rates, the rates themselves are spatially clustered for A10, C, and C10 ($I=0.36$, $0.45$, and $0.42$; all $p_{\mathrm{sim}}=0.001$).
Adjusting for $S$ reduces this autocorrelation but does not completely remove it ($I=0.27$, $0.35$, and $0.34$; all $p_{\mathrm{sim}}=0.001$).
Adding the nine community intensities $I_c$ reduces Moran's $I$ to $0.19$, $0.03$, and $0.15$ ($p=0.002$, $0.214$, and $0.005$), leaving no evidence of residual spatial dependence for C.
Among residents aged $\geq 60$, residual clustering remains high after $S$ ($I=0.54$, $0.35$, and $0.43$), while $S{+}I$ lowers these values ($I=0.19$, $0.11$, and $0.25$).
Sociodemographic covariates alone are therefore insufficient to explain spatial correlation between neighboring municipalities; community-level indicators absorb spatial structure that demographic factors leave unexplained.

\begin{figure}[h!]
\centering
\includegraphics[width=\textwidth]{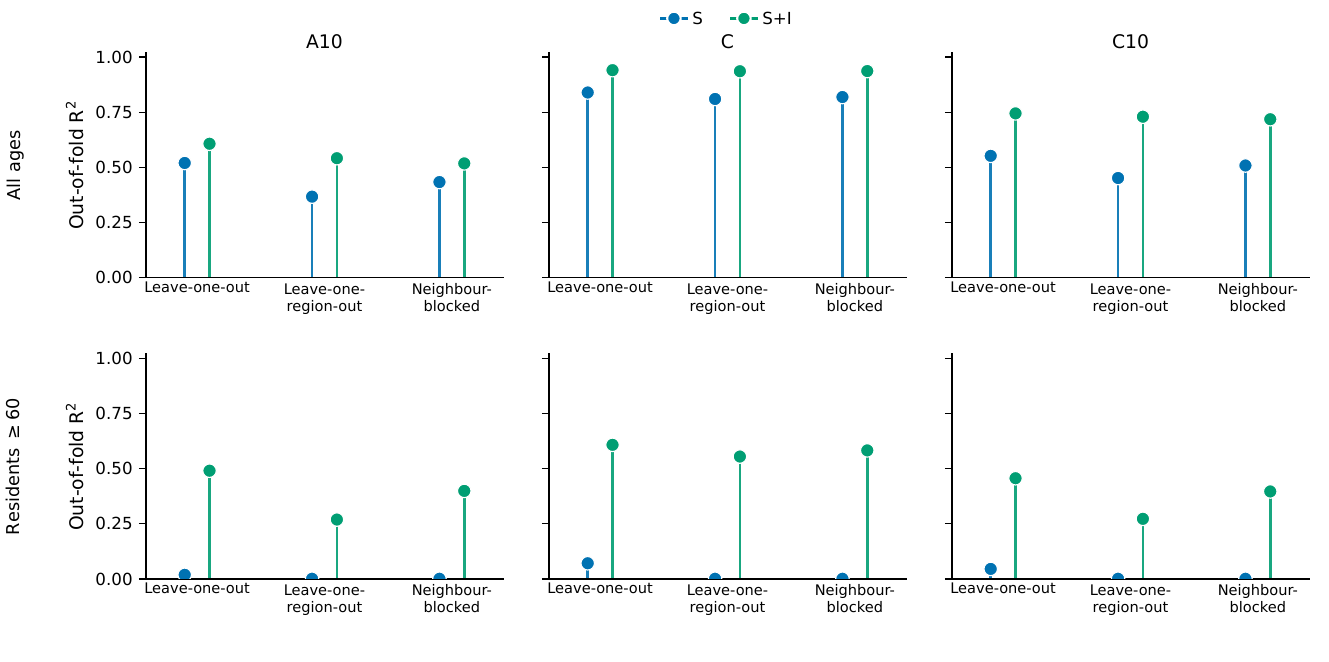}
\caption{\textbf{Out-of-fold predictive performance across spatial cross-validation schemes.}
    Out-of-fold coefficient of determination ($R^2$) from Ridge regression models predicting municipal prescription rates using sociodemographic covariates alone ($S$) and including the nine community intensities ($S + I$).
    From left to right, panels show models for antidiabetic (A10), cardiovascular (C), and lipid-modifying (C10) prescription rates, for age-aggregated rates (top) and residents aged $\geq 60$ (bottom).
    Within each panel, we compare ordinary leave-one-out, neighbor-blocked leave-one-out (holding out municipality $i$ and its five nearest neighbors while predicting only $i$), and leave-one-region-out schemes; bars report models with $S$ alone versus $S + I$.
    }
\label{sm_spatial_cv}
\end{figure}

Ordinary leave-one-out Ridge can leak spatial information if neighboring municipalities stay in the training set.
We therefore repeated the leave-one-out Ridge models of the main analysis with leave-one-region-out and with neighbor-blocked leave-one-out, which holds out municipality $i$ and its five nearest neighbors and predicts only $i$.
Figure \ref{sm_spatial_cv} reports out-of-fold $R^2$ for $S$ and $S{+}I$, showing that the increment after $S$ remains in both cases.
For age-aggregated rates, the residual shares after $S$ (A10/C/C10) are $18/63/43\%$ (leave-one-out), $15/65/43\%$ (neighbor-blocked), and $28/67/51\%$ (leave-one-region-out), while among residents aged $\geq 60$, $S$ stays near zero out of fold whereas $S{+}I$ remains large (neighbor-blocked residual shares: $45/60/45\%$).
Taken together, these results demonstrate that the predictive power of community intensities is robust to spatial dependency, absorbing municipal-level spatial structure without relying on local information leakage.
\pagebreak[2]
\clearpage
\section*{Supplementary Materials, Section 5: Structural Validation of Food Mapping and Processing Scores}
To quantify the performance of individual mapping tiers, we benchmarked each step independently against Open Food Facts NOVA classifications \cite{monteiro_ultra-processed_2019}.
Direct FoodEx2 linkage \cite{european_food_safety_authority_efsa_food_2015} ($n = 783$) achieved the highest fidelity ($69.1\%$ overall accuracy, UPF $F_1 = 0.846$, and $79.7\%$ on NOVA 1 vs. 4), but covered only approximately one-quarter of the retail catalog.
Unconstrained string matching alone reduced performance to $61.7\%$ overall accuracy ($F_1 = 0.763$; $69.2\%$ on NOVA 1 vs. 4), primarily due to token collisions across processing states (e.g., canned fish matching raw fish, chocolate dairy snacks matching buttermilk via the shared token ``milk'', or snack bars matching peanut butter).
When mapped Frida foods and local on-pack nutrient profiles were compositionally close (top distance tercile), categorical NOVA agreement reached $87.9\%$ with a Pearson correlation of $r = 0.70$ on the continuous processing score, confirming that generic Frida proxies preserve the product-level processing signal.

\begin{figure}[h!]
\centering
\includegraphics[width=\textwidth]{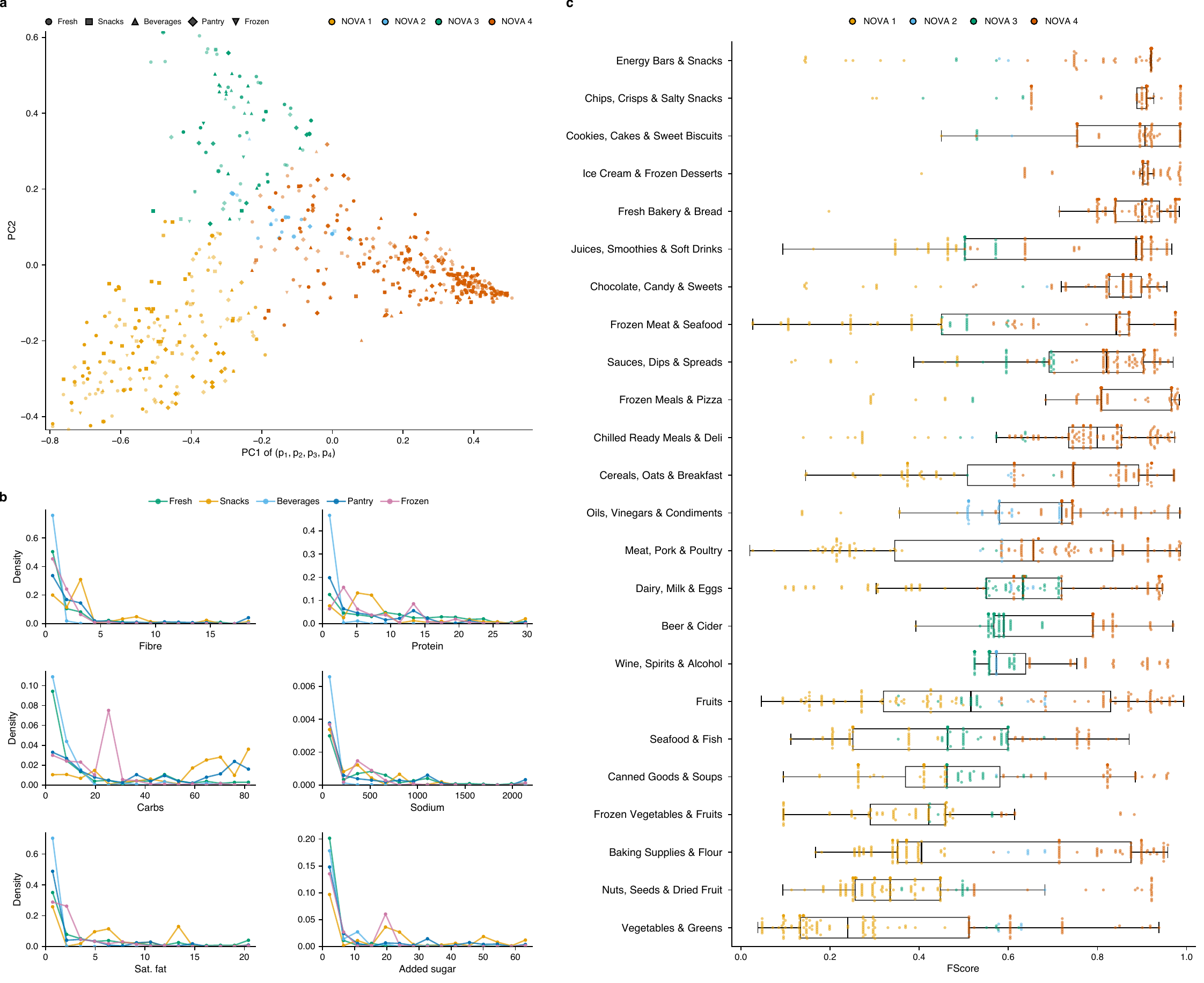}
\caption{\textbf{Nutritional composition and industrial processing landscape of the mapped retail catalog.}
\textbf{a)} Principal component analysis (PCA) of FoodProX predicted class probabilities $(p_1, p_2, p_3, p_4)$ across mapped retail products ($n = 8{,}672$). Points are colored by assigned FoodProX NOVA class (1: unprocessed/minimally processed to 4: ultra-processed), with marker shapes denoting catalog macro-families (Fresh Food, Snacks \& Sweets, Beverages, Pantry \& Packaged Food, Frozen Foods). The primary axis separates minimally processed foods (NOVA 1, left, orange) from ultra-processed formulations (NOVA 4, right, red) along a continuous processing gradient.
\textbf{b)} Empirical distributions of key nutrient densities (dietary fiber, protein, available carbohydrates, sodium, saturated fat, and added sugar in g or mg per 100\,g edible portion of the mapped Frida food) across the five catalog macro-families (colors). Fresh foods exhibit high protein and fiber densities, whereas snacks and sweets show pronounced skews toward added sugar and saturated fat.
\textbf{c)} Store subcategories ($n \ge 50$ mapped products) ranked by median continuous processing score ($F\text{Score} \in [0, 1]$).
The ranking spans from raw vegetables and dried nuts to industrial bakery goods, cookies, and crisps, with tightly constrained dispersion in confectioneries contrasting with wide heterogeneity in breakfast cereals and dairy.
}
\label{sm_fscore}
\end{figure}

After mapping, each retail product inherits the complete Frida nutrient profile, the FoodProX class probability simplex $(p_1, p_2, p_3, p_4)$, and the continuous processing metric $F\text{Score}$ \cite{poulsen_dataset_2026,menichetti_machine_2023} ($n = 8{,}672$ food products mapped to $729$ distinct Frida reference foods, excluding non-food lines).
Principal component analysis of $(p_1, p_2, p_3, p_4)$ reveals a clear separation between minimally processed items (NOVA 1) and ultra-processed formulations (NOVA 4), forming two distinct clusters at opposite ends of the first principal component (see Figure \ref{sm_fscore}\textbf{a}).
Fresh commodities (vegetables, fresh meat, produce, seafood) concentrate at the minimally processed pole, whereas confectioneries, savory snacks, and industrial bakery products anchor the ultra-processed extreme; frozen and packaged pantry goods bridge the intermediate space.
Nutrient distributions systematically reflect these processing differences (see Figure \ref{sm_fscore}\textbf{b}), with fresh categories supplying the bulk of dietary fiber and protein, while snacks and sweets exhibit marked shifts toward saturated fats and added sugars.
Ranking store subcategories by median $F\text{Score}$ establishes a coherent industrial processing gradient (see Figure \ref{sm_fscore}\textbf{c}) ranging from raw produce ($F\text{Score} \approx 0.24$) to industrial chips and snack bars ($F\text{Score} > 0.90$).
While categories such as potato chips and commercial bakery items display tight within-group uniformity, others -- notably breakfast cereals and dairy -- exhibit broad compositional heterogeneity.
This empirical ordering aligns with US grocery benchmarks \cite{ravandi_prevalence_2025}, with particularly strong concordance among the highest-ranked snack, bakery, sweet, and frozen-dessert categories and broadly consistent ordering of staple, dairy, meat, and produce categories toward the middle and lower ends, confirming the structural validity of the crosswalk.

\pagebreak[2]
\clearpage
\section*{Supplementary Materials, Section 6: Basket Compositional Context of Co-Purchasing Ties}
To determine whether the validated co-purchasing backbone is supported by mixed baskets or by mutually exclusive transaction types (such as separate fresh-food versus ultra-processed receipts), we evaluated the compositional context of validated product pairs.
All evaluations are conducted on the national Bipartite Configuration Model skeleton used in the main text \cite{saracco_inferring_2017} (see Methods and Supplementary Materials, Section 3).
We avoid splitting receipts into boolean subsets (such as checkouts with versus without NOVA-4 items), reconstructing separate null models on sub-graphs, or computing likelihood differences ($\Delta \log L$) on arbitrary transaction subsets, as such stratified constructions conflate sample size $N$, degree sequences, and basket sizes $k$, confounding genuine topological association with marginal size distributions.

Let $I_r$ denote the set of mapped brand $\times$ micro-category products in receipt $r$, $k_r = |I_r|$ the basket size, $n_{4,r}$ the count of ultra-processed items (NOVA 4 via FoodProX \cite{menichetti_machine_2023,monteiro_ultra-processed_2019}), and $p_4(r) = n_{4,r} / k_r$ the local UPF share.
For any validated skeleton edge $e = (i, j)$, let $R_e = \{r : i, j \in I_r, \, k_r \ge 3\}$ denote the set of supporting receipts containing both items with a minimum checkout size of three items.
We define the tie carriage $\gamma_e$ as the expected UPF share among baskets supporting edge $e$, $\gamma_e = \mathbb{E}\left[ p_4(r) \mid r \in R_e \right]$.
By construction, $\gamma_e$ is directly comparable to aggregate purchasing baselines, where the item-weighted national UPF share is $p_4 = 0.52$ and the mean UPF share across mixed baskets is $0.49$.
We compute national values by pooling municipal receipts weighted by supporting transaction volume $N_{3,e} = |R_e|$ rather than equal municipal weights, retaining edges with $N_{3,e} \ge 50$ and both endpoints FoodProX-mapped to ensure estimation stability ($82.7\%$ of total edges).

Across 48.7 million food receipts, $47.9\%$ contain both a NOVA-1 and a NOVA-4 item.
These mixed carts account for $76.6\%$ of all mapped retail volume, and $83.3\%$ of receipts containing at least one NOVA-1 item also contain at least one NOVA-4 item.
Because the endpoints of edge $e = (i, j)$ enter $p_4(r)$ directly, the class distributions of $\gamma_e$ are partially separated by arithmetic constraints: $p_4 \ge 0$ for 1–1 pairs, $p_4 \ge 1/k_r$ for 1–4 pairs, and $p_4 \ge 2/k_r$ for 4–4 pairs.
The substantive result is that minimally processed pairs are not confined to fresh-food baskets, with $\bar{\gamma}_{11}=0.353$ (95\% bootstrap CI: $[0.342, 0.364]$), a value that sits well within mixed carts rather than isolated trips (see Supplementary Figure \ref{sm_fig_upf_context}\textbf{a}).
In contrast, mixed and ultra-processed pairs yield $\bar{\gamma}_{14} = 0.452$ and $\bar{\gamma}_{44} = 0.592$, showing that UPF--UPF ties concentrate in already UPF-dense baskets.

To test whether tie carriage reflects genuine dyadic structure or simply the marginal composition of the linked products, we computed a leave-two-out contrast $\Delta_e$ against a municipality-specific exact-$k$ baseline ($\bar{\Delta}_{11} = -0.123$, $\bar{\Delta}_{44} = +0.117$).
Because $\Delta_e$ is almost entirely accounted for by an additive combination of marginal product means [$R^2(\Delta_e \sim \frac{u_i + u_j}{2}) = 0.95$, with Spearman rank correlations to mean basket size $\bar{k}$ of $0.00$, $-0.37$, and $-0.67$ within 1--1, 1--4, and 4--4 pairs], it captures node-level composition rather than residual tie context.
Projecting carriage directly to the product level via the incident $N_{3,e}$-weighted mean, $\tilde{\gamma}_i = \sum_{e \in \partial i} N_{3,e}\gamma_e / \sum_{e \in \partial i} N_{3,e}$, shows that core minimally processed staples cluster tightly near $\tilde{\gamma}_i \approx 0.38$--$0.40$, whereas branded carbonated soft drinks reach $0.68$ (Supplementary Figure \ref{sm_fig_upf_context}\textbf{b}).
Spatially, municipal mean $\bar{\gamma}$ displays a discrete level shift between the urban core and the rest of the country rather than a continuous gradient (Supplementary Figure \ref{sm_fig_upf_context}\textbf{c}).
Relative to the urban core, municipal $\bar{\gamma}_{11}$ is approximately $9$--$10\%$ higher in Near and Mid--far municipalities, whereas $\bar{\gamma}_{44}$ remains within about $2\%$ of that urban baseline.
This divergence represents a baseline offset defined by the urban core rather than a continuous gradient along the spectral axis, with neither class showing a significant rank correlation with spectral distance ($\rho = 0.17$ for 1--1; $\rho = 0.02$ for 4--4; $p>0.05$).
\begin{figure}[!ht]
    \centering
    \includegraphics[width=\textwidth]{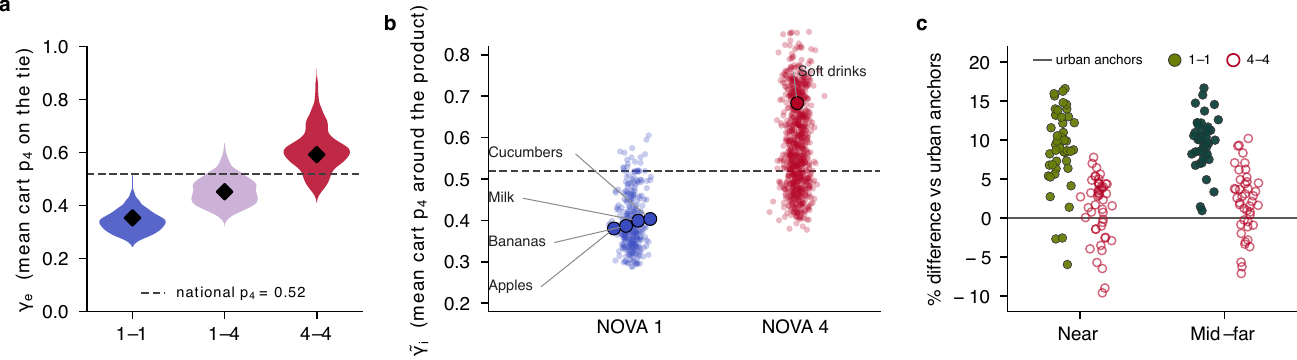}
    \caption{\textbf{Basket compositional context of BiCM-validated co-purchasing ties.}
    \textbf{a)} Edge-level carriage $\gamma_e$, defined as the mean basket UPF share $p_4$ across transactions supporting each edge.
    Distributions are shown across processing pairs: minimally processed (NOVA 1--1, blue), mixed (NOVA 1--4, purple), and ultra-processed (NOVA 4--4, red).
    Black diamonds indicate $N_{3,e}$-weighted class means ($\bar{\gamma}_{11} = 0.353$, $\bar{\gamma}_{14} = 0.452$, $\bar{\gamma}_{44} = 0.592$).
    The horizontal dashed line marks the item-weighted national baseline ($p_4 = 0.52$).
    \textbf{b)} Product-level carriage $\tilde{\gamma}_i$, computed as the $N_{3,e}$-weighted mean of incident $\gamma_e$ values across NOVA-1 and NOVA-4 products.
    Common Staples (apples, bananas, milk, cucumbers) sit near $0.38$--$0.40$, while ultra-processed formulations (e.g., carbonated soft drinks) reach $0.68$.
    \textbf{c)} Percentage difference in municipal mean tie carriage $\bar{\gamma}$ relative to the urban core (horizontal line at zero), grouped into Near and Mid--far municipalities (with filled circles NOVA 1--1 and open circles NOVA 4--4).
    The peripheral shift reflects a discrete level shift in 1--1 ties ($+9$ to $+10\%$) rather than a continuous spectral gradient.
    }
    \label{sm_fig_upf_context}
\end{figure}
\end{document}